\documentstyle[epsf,bookmath]{lamuphys}

\makeatletter
\let\chapter\hid@chapter
\makeatother

\begin{document}
\pagenumbering{arabic}

%
%       M. Eschrig
%
\title{
 Electromagnetic Response of a Pancake Vortex
 in Layered Superconductors
}
\author{Matthias Eschrig}
\institute{Physikalisches Institut, Universit\"at Bayreuth,
D-95440 Bayreuth, Germany}
\titlerunning{Electromagnetic Response of a Pancake Vortex  ...}
\maketitle
\begin{abstract}
We use the quasiclassical theory of superconductivity to
calculate the response of the currents in  the core of a 
vortex to an alternating electric field. We consider relatively
clean
superconductors with a mean free path, $\ell$, 
%of the order of
%the coherence length, $\xi_0$, or larger.
larger than the coherence length, $\xi_0$.
The current response of the core is dominated by the 
bound states of Caroli, de Gennes and Matricon,
 and differs   significantly from the response obtained from
the model of a ``normal core'' introduced by Bardeen and Stephen.
The model of Bardeen and Stephen describes the limit of a dirty core
($\ell \ll \xi_0$), and fails for a clean superconductor. 
The response
of the bound states  includes 
non-dissipative
acceleration of the charge carriers
as well as dissipative
currents. 
Dissipation dominates at low frequencies ($\hbar\omega\la
.5\Delta$), and ${\cal R}e\sigma(\omega)$ greatly exceeds the
normal state conductivity. 
These dissipative currents have a 
dipolar structure superimposed on an oscillating
supercurrent in the direction of the electric field.
We attribute the anomalous dissipation 
to the zero-energy bound state
broadened by lifetime effects.
\footnote{This work was
done in collaboration with Dierk Rainer and James A. Sauls.}
\end{abstract}
\section{Introduction}

An electric field has two principal effects on a superconductor. 
It changes the supercurrents by accelerating the superfluid condensate, 
and it generates  dissipation by exciting ``normal'' quasiparticles. This two-fluid picture
 of a {\it condensate} and {\it normal excitations} is clearly reflected in the 
optical conductivity of  standard s-wave superconductors in the
Meissner state 
(\cite{Mattis58}).
The conductivity 
has a superfluid part, 
\begin{equation}\label{rhos}
\sigma_s(\omega)
%= i{e^2n_s\over m}{1\over \omega+i0}
= {e^2n_s\over m}\left(\delta(\omega)+iP{1\over\omega}\right)\enspace ,
\end{equation}
and a dissipative normal part. At $T=0$ dissipation starts at the 
threshold frequency for creating quasiparticles,  
 $\omega=2\Delta$. At finite temperature the coupling to 
thermally excited quasiparticles leads  to 
 dissipation at all frequencies. 
The  two-fluid picture  no longer holds for type-II superconductors
in the vortex state.  The response to an electric field consists
of contributions from  
gapless\footnote{We ignore  the ``mini-gap'' of size
$\Delta^2/E_f$ predicted by \cite{Caroli64}.}
vortex core  excitations (\cite{Caroli64}), and
contributions from excitations outside the cores, 
where the quasiparticle spectrum shows  the bulk  gap.
For high-$\kappa$ superconductors  
the electric response at low frequencies of the region outside the
core can be 
described 
 very well by the London equations (\cite{London50}), i.e. by 
two-fluid electrodynamics. The 
 response of the  core  is more complex. In the 
 traditional model of a 
``normal core'' (Bardeen and Stephen, 1965) the 
conductivity in the core 
is that of the normal state, $\sigma_{core}=\sigma_n$. 
The Bardeen-Stephen model 
is a plausible approximation for dirty superconductors with a 
mean free path ($\ell$) much shorter than the size  
of the core ($\ell\ll\xi_0$).
In this limit the vortex core  excitations of \cite{Caroli64} may be considered as a continuum of
normal excitations.

In 1969 Bardeen et al. published  a detailed discussion of
the bound states in the core of a vortex, and argued  that the 
bound states contribute significantly to the 
circulating  supercurrents  in the vortex core.
This effect  was confirmed by recent self-consistent
calculations of free and pinned vortices in 
clean and medium dirty superconductors (\cite{Rainer96}). 
The authors show that all currents in the core
(circulating currents as well 
as  superimposed transport currents) are 
carried predominantly  by the
bound states. This means that the model of a normal core 
needs to be modified for relatively   
clean  ($\ell\ga\xi_0$) superconductors.
Furthermore the response of the bound states 
is expected to show  dissipative  as well as  superfluid features. 

In this article we  present calculations of 
 the linear response of the current in the core of an isolated
 vortex to an {\it a.c.} electric field with  frequency below the 
bulk threshold frequency $2\Delta$. In section 2 we
describe the model as well as  the basic theoretical concepts and equations
of the quasiclassical linear response theory of inhomogeneous
superconductors. We discuss the importance of
vertex corrections, charge conservation, and 
 the local charge-neutrality condition. 
Self-consistent calculations of the structure and excitation
spectrum of a  vortex in equilibrium 
are presented in section 3. The analysis and  discussion 
emphasizes of the 
effects of impurities on the bound states in the core. Our
numerical methods are based on a formulation of the
quasiclassical response theory in terms of ``distribution
functions'', which are introduced in section 4. In sections
5 and 6 we present and discuss our results for the current 
response to a small {\it a.c.}  electric field.

\section{Quasiclassical theory of the electromagnetic response}

Our calculation of the response of a vortex to an {\it a.c.} 
electric field is based on 
the quasiclassical theory of Fermi liquid superconductivity. 
This theory describes phenomena on length
scales large compared to  the microscopic scales (Bohr radius, 
lattice constant, $k_f^{-1}$, Thomas-Fermi
screening length, etc.) and frequencies small compared to the microscopic
scales (Fermi energy, plasma frequency, conduction band width, etc.). 
\footnote{We set in this article $\hbar=k_B=1$. The charge of an 
electron is $e < 0$\ . } The {\it a.c.} frequencies we consider 
are of the order of the superconducting energy gap, 
$\mid\Delta\mid\sim T_c$, or 
smaller, and the length scales of interest are the coherence 
length, $\xi_0= \mbox{v}_f/2\pi T_c$, and the penetration 
depth, $\lambda$. Hence, our theory  requires the
conditions $k_f\xi_0 \gg 1$, and $T_c/E_f \ll 1$, where 
the Fermi wavelength, $k_f^{-1}$,
and the Fermi energy, $E_f$, stand for typical microscopic 
length and energy scales.

A convenient formulation of the quasiclassical theory for our
purposes is in terms of the quasiclassical Nambu-Keldysh propagator 
$\check{g}(\vec{p}_f,\vec{R};\epsilon,t)$, which is a $4\times
4$-matrix in Nambu-Keldysh space, and a function of position
$\vec{R}$, time $t$, energy $\epsilon$, and Fermi momentum $\vec{p}_f$. 
%\footnote{For details of our notation see  \cite{Rainer97}.}
We consider a superconductor
with random atomic size impurities
in a static magnetic field,
and an externally applied {\it a.c.} electric field, 
$\vec{E}=-{1\over c}\partial_t\delta\vec{A}$. 
In a compact notation the  transport equation for this system and  the
normalization condition read 
\footnote{
The commutator $[\check A,\check B]_\otimes $ is given by
$\check A \otimes \check B- \check B \otimes \check A$, were the
noncommutative $\otimes $-product is defined by 
$\check A \otimes \check B(\epsilon,t)=\exp^{\frac{i}{2} (\partial_\epsilon^A \partial_t^B-\partial_t^A \partial_\epsilon^B) } \check A(\epsilon,t)\check  B(\epsilon,t)$ .}
\begin{equation}\label{transpqcl}
\bigl[(\epsilon+{e\over c}\vec{v}_f\cdot\vec{A})
\check\tau_3-\check\Delta_{mf}-\check\sigma_{i}-\delta\check{v}\,,\check
g\bigr]_{\otimes} +i\vec{v}_f\cdot\vec\nabla\check g\\ = \ 0\enspace ,
\end{equation}
\begin{equation}\label{normalize}
\check g\otimes \check g=-\pi^2\check{1}\enspace ,
\end{equation}
where $\vec{A}(\vec R)$ is the vector potential of the static
magnetic field,
$\vec{B} =\vec{\nabla}\times\vec{A}$,
$\check\Delta_{mf}(\vec{p}_f,\vec{R};t)$ the mean-field 
order parameter
matrix, and $\check\sigma_{i}(\vec{p}_f,\vec{R};\epsilon,t)$ 
is the impurity self-energy. 
The perturbation  $\delta\check{v}(\vec{p}_f,\vec{R};t)$ includes
the  external electric field and
the field of  the charge fluctuations, $\delta\rho(\vec{R};t)$,
induced by the external field.
For convenience we describe the external electric field by a vector 
potential $\delta\vec{A}(\vec R;t)$ and the induced 
electric field by 
the electro-chemical potential $\delta\varphi(\vec{R};t)$. Hence
in the Nambu-Keldysh matrix notation
the perturbation has the form,  
\begin{equation}\label{perturbation}
\delta\check{v} =
-{e\over c}\vec{v}_f\cdot\delta\vec{A}(\vec R;t)\check{\tau}_3
+e\delta\varphi(\vec{R};t)\check{1}\enspace , 
\end{equation}
and is assumed to be sufficiently  small so that it 
can be treated in linear response theory.

Equations (\ref{transpqcl}) and (\ref{normalize}) 
must be supplemented by 
self-consistency equations for the order parameter and the impurity
self-energy. We use  the  weak-coupling gap equations,
\begin{equation}\label{gapequation}
\hat\Delta^{R,A}_{mf}(\vec{p}_f,\vec{R};t)=
\int_{-\epsilon_c}^{+\epsilon_c}{d\epsilon\over 4\pi i}
\big<V(\vec{p}_f,\vec{p}_f^{\prime})
\hat f^K(\vec{p}_f^{\prime},\vec{R};\epsilon,t)\big>
\enspace ,
\end{equation}
\begin{equation}\label{Kgapequation}
\hat\Delta^{K}_{mf}(\vec{p}_f,\vec{R};t)=0\enspace ,
\end{equation}
and the impurity self-energy in Born approximation with isotropic
scattering,
\begin{equation}\label{born}
\check\sigma_{i}(\vec{R};\epsilon,t)=
{1\over 2\pi\tau}
\big<\check{g}(\vec{p}_f^{\prime},\vec{R};\epsilon,t)\big>
\enspace ,
\end{equation}
where 
$\hat{f}^K$ is the off-diagonal part of the $2\times 2$ Nambu
matrix $\hat g^K$, and 
the Fermi surface average is defined by
\begin{equation}\label{average}
\big<\ldots\big>\, =\ 
{1\over N_f}\int{d^2\vec{p}_f^{\prime}\over (2\pi)^3\mid\!\vec{v}_f
^{\prime}\!\mid}\ \ldots 
\enspace .
\end{equation}
The materials parameters that enter the self-consistency equations are 
the pairing interaction, $V(\vec{p}_f,\vec{p}_f^{\prime})$,  the 
impurity scattering lifetime, $\tau$, in addition to the Fermi surface data 
$\vec{p}_f$ (Fermi surface), $\vec{v}_f$ (Fermi velocity), and 
$N_f=\int{d^2\vec{p}_f\over (2\pi)^3\mid\vec{v}_f\mid}$.

In the linear response approximation  one splits the  propagator and
the self-energies into an unperturbed part and a term  of 
first order in
the perturbation,
\begin{equation}\label{linresp}
\check{g}=\check{g}_0+\delta\check{g}\,, \ \ 
\check{\Delta}_{mf}=\check{\Delta}_{mf0} +\delta\check{\Delta}_{mf}\,, \ \
\check{\sigma}_{i}=\check{\sigma}_{0}+\delta\check{\sigma}_{i}
\enspace ,
\end{equation}
and expands the transport equation and normalization condition 
through first order. In 0$^{th}$ order we obtain
\begin{equation}\label{transpqcl0}
\bigl[(\epsilon+{e\over c}\vec{v}_f\cdot\vec{A})
\check\tau_3-\check\Delta_{mf0}-\check\sigma_{0}\,,\check
g_0\bigr]_{\otimes} +i\vec{v}_f\cdot\vec\nabla\check g_0\\ = \ 0\enspace ,
\end{equation}
\begin{equation}\label{normalize0}
\check{g}_0\otimes\check{g}_0=-\pi^2\check{1}\enspace ,
\end{equation}
and in 1$^{st}$ order
\begin{equation}\label{transpqcl1}
\bigl[(\epsilon+{e\over c}\vec{v}_f\cdot\vec{A})
\check\tau_3-\check\Delta_{mf0}-\check\sigma_{0}\,,\delta\check
g\bigr]_{\otimes} +i\vec{v}_f\cdot\vec\nabla\delta\check g\\ = \
\bigl[
\delta\check{\Delta}_{mf}+\delta\check{\sigma}_{i}+\delta\check{v}\,,\
\check{g}_0]_{\otimes}\enspace ,
\end{equation}
\begin{equation}\label{normalize1}
\check{g}_0\otimes\delta\check{g}
+\delta\check{g}\otimes\check{g}_0=0\enspace .
\end{equation}
In order to close this system of equations one has to supplement 
the transport and normalization equations with the
 self-consistency equations of 0$^{th}$  and 1$^{st}$ order:
\begin{equation}\label{lgapequation0}
\hat\Delta^{R,A}_{mf0}(\vec{p}_f,\vec{R})=
\int_{-\epsilon_c}^{+\epsilon_c}{d\epsilon\over 4\pi i}
\big<V(\vec{p}_f,\vec{p}_f^{\prime})
\hat f^K_0(\vec{p}_f^{\prime},\vec{R};\epsilon)\big>
,\enspace  \hat\Delta^K_{mf0}=0\enspace ,
\end{equation}
\begin{equation}\label{lgapequation1}
\delta\hat\Delta^{R,A}_{mf}(\vec{p}_f,\vec{R};t)=
\int_{-\epsilon_c}^{+\epsilon_c}{d\epsilon\over 4\pi i}
\big<V(\vec{p}_f,\vec{p}_f^{\prime})
\delta\hat f^K(\vec{p}_f^{\prime},\vec{R};\epsilon,t)\big>,
\enspace 
\delta\hat\Delta^{K}_{mf}=0\enspace ,
\end{equation}
and 
\begin{equation}\label{respborn0}
\check\sigma_{0}(\vec{R};\epsilon)=
{1\over 2\pi\tau}
\big<\check{g}_0(\vec{p}_f^{\prime},\vec{R};\epsilon)\big>
\enspace ,
\end{equation}
\begin{equation}\label{respborn1}
\delta\check\sigma_{i}(\vec{R};\epsilon,t)=
{1\over 2\pi\tau}
\big<\delta\check{g}(\vec{p}_f^{\prime},\vec{R};\epsilon,t)\big>
\enspace .
\end{equation}

Finally, the electro-chemical potential, $\delta\varphi$, is
determined by the condition of local charge neutrality 
(\cite{Gorkov75}, \cite{Artemenko79}).
This condition follows from the expansion of charge density 
to leading order in the quasiclassical expansion parameters. One
obtains $\delta\rho(\vec{R};t)=0$, i.e.
\begin{equation}\label{localneutral}
-2e^2N_f\delta\varphi(\vec{R};t)+eN_f
%\int_{-\epsilon_c}^{+\epsilon_c}{d\epsilon\over
\int{d\epsilon\over
4\pi i}\big<\mbox{Tr}
\delta\hat g^K(\vec{p}_f^{\prime},\vec{R};\epsilon,t)\big>
= \,0
\enspace .
\end{equation}
The self-consistency equations 
(\ref{lgapequation0})-(\ref{respborn1})
are of vital  importance in the context of this paper. Equations 
(\ref{lgapequation1}) and (\ref{respborn1}) for the response of the
quasiclassical self-energies are  equivalent  to  
{\em vertex corrections} in the Green's function response theory. 
They  guarantee that the quasiclassical theory does not
violate fundamental conservation laws.
 In particular,  (\ref{respborn0}) and (\ref{respborn1})
imply 
charge conservation in scattering processes,  whereas
(\ref{lgapequation0}) and  (\ref{lgapequation1})
imply charge conservation in a particle-hole conversion process.
Any charge which is lost   or gained in a
 particle-hole conversion process is balanced  by the corresponding
 gain or loss of  condensate charge. It is the coupled
 quasiparticle dynamics and collective condensate dynamics
 which conserves charge in superconductors. 
 Neglect of the dynamics of either component,  or use of a non-conserving
 approximation for the coupling of quasiparticles and
 collective degrees of freedom
 leads to unphysical results. 
 Condition (\ref{localneutral}) 
 is a consequence of the long-range of the  
 Coulomb repulsion. The Coulomb energy of a charged 
 region of size
 $\xi_0^3$ and typical charge density $eN_f\Delta$ is  
 $\ \sim e^2N_f^2\Delta^2\xi_0^5$, which should be compared
 with the condensation energy $\sim N_f\Delta^2\xi_0^3$. 
 Thus,  the cost in Coulomb energy is a factor $(E_f/\Delta)^2$
 larger than the condensation energy.  This  leads to a strong 
 suppression of charge
 fluctuations, and  the condition  of local charge 
 neutrality holds to very  good accuracy
 for superconducting phenomena.

Equations (\ref{transpqcl0})-(\ref{localneutral}) constitute
a complete set of equations for calculating the 
electromagnetic response of a vortex. The 
structure of a vortex in equilibrium is
obtained from (\ref{transpqcl0}), (\ref{normalize0}), 
(\ref{lgapequation0}) and
(\ref{respborn0}), and the linear response
of the vortex to the perturbation
$\delta\vec{A}(\vec{R};t)$ follows from (\ref{transpqcl1}),
(\ref{normalize1}),
(\ref{lgapequation1}), (\ref{respborn1}) and (\ref{localneutral}).
The currents   induced by
$\delta\vec{A}(\vec{R};t)$ can then be calculated directly from the 
Keldysh propagator $\delta\hat g^K$ via 
\begin{equation}\label{current1}
\delta\vec{j}(\vec{R};t)= eN_f
\int {d\epsilon\over
4\pi i}\big<\vec{v}_f(\vec{p}_f^{\prime})\mbox{Tr}\left(\hat\tau_3
\delta\hat g^K(\vec{p}_f^{\prime},\vec{R};\epsilon,t)
\right)\big>
\enspace .
\end{equation}

\section{Structure and Spectrum of a Pancake Vortex
in Equilibrium}

We consider an isolated pancake vortex in a strongly
anisotropic, layered superconductor, and  
model the Fermi surface 
of these systems   by a cylinder of radius $p_f$, and a 
Fermi velocity of   constant magnitude, $v_f$, 
 along the layers.
We also
assume  isotropic (s-wave) pairing and isotropic
impurity scattering. Thus, the materials 
parameters of the   model superconductor are:
$T_c$, $v_f$, the 2D density of states $N_f=p_f/2\pi v_f$, 
and the mean free path
$\ell= v_f\tau$. The model superconductor is 
type II with a large 
Ginzburg-Landau parameter $\kappa \ga 100$, so the magnetic
field is  to  good approximation constant in the region of
the vortex core.  

We first present results for the equilibrium
vortex. The order parameter, $\Delta_0(\vec{R})$, is
calculated by solving  the transport
equation (\ref{transpqcl0}), the gap equation
(\ref{lgapequation0}), and the self-energy equation
(\ref{respborn0}) self-consistently. 
These calculations are done at Matsubara
energies ($\epsilon \rightarrow i\epsilon_n = i(2n+1)\pi T$).
Details of the numerical schemes for solving the transport
equation self-consistently are given elsewhere
(\cite{Eschrig97}). 
Charge conservation for the equilibrium 
vortex follows from the circular  symmetry of the currents. 
Nevertheless, self-consistency of the equilibrium vortex
is important; 
the equilibrium self-energies
($\check\Delta_0$,  $\check\sigma_0$)
and propagators ($\check g_0$) are input quantities in
the transport equation (\ref{transpqcl1}) for the linear
response. Charge conservation in linear response 
is non-trivial and requires self-consistency of both the
equilibrium solution and the solution in first order in the
perturbation.

\begin{figure}[t]
%%%%%%%%%%%%%%%%%%%   F I G U R E   %%%%%%%%%%%%%%%%%%%%
\begin{minipage}[t]{6.0cm}
\centerline{
\epsfxsize6.0cm
\epsffile{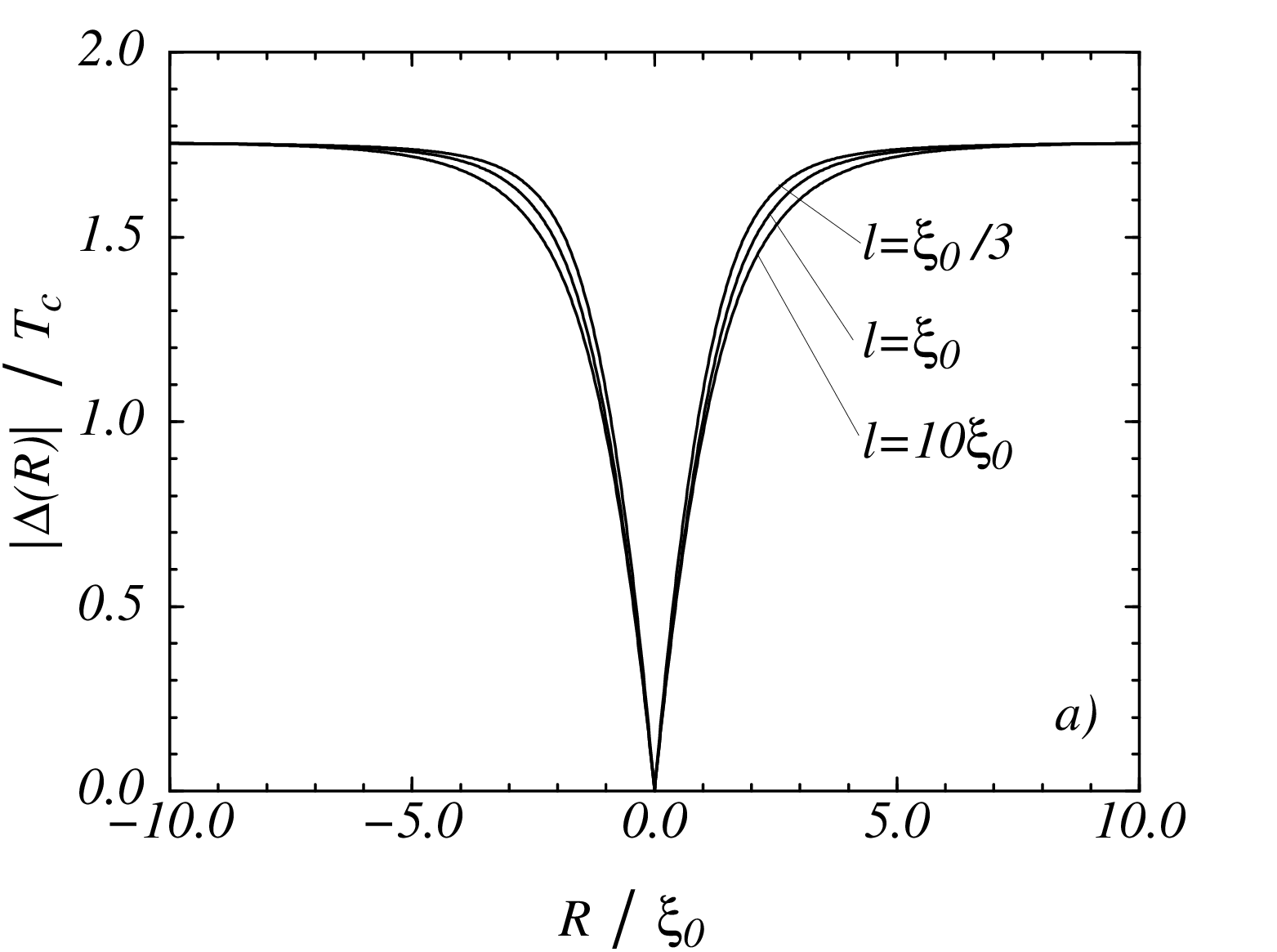}
}
%%%%%%%%%%%%%%%%%%%%%%%%%%%%%%%%%%%%%%%%%%%%%%%%%%%%%%%%
\end{minipage}
\hfill
\begin{minipage}[t]{6.0cm}
%%%%%%%%%%%%%%%%%%%   F I G U R E   %%%%%%%%%%%%%%%%%%%%
\centerline{
\epsfxsize6.0cm
\epsffile{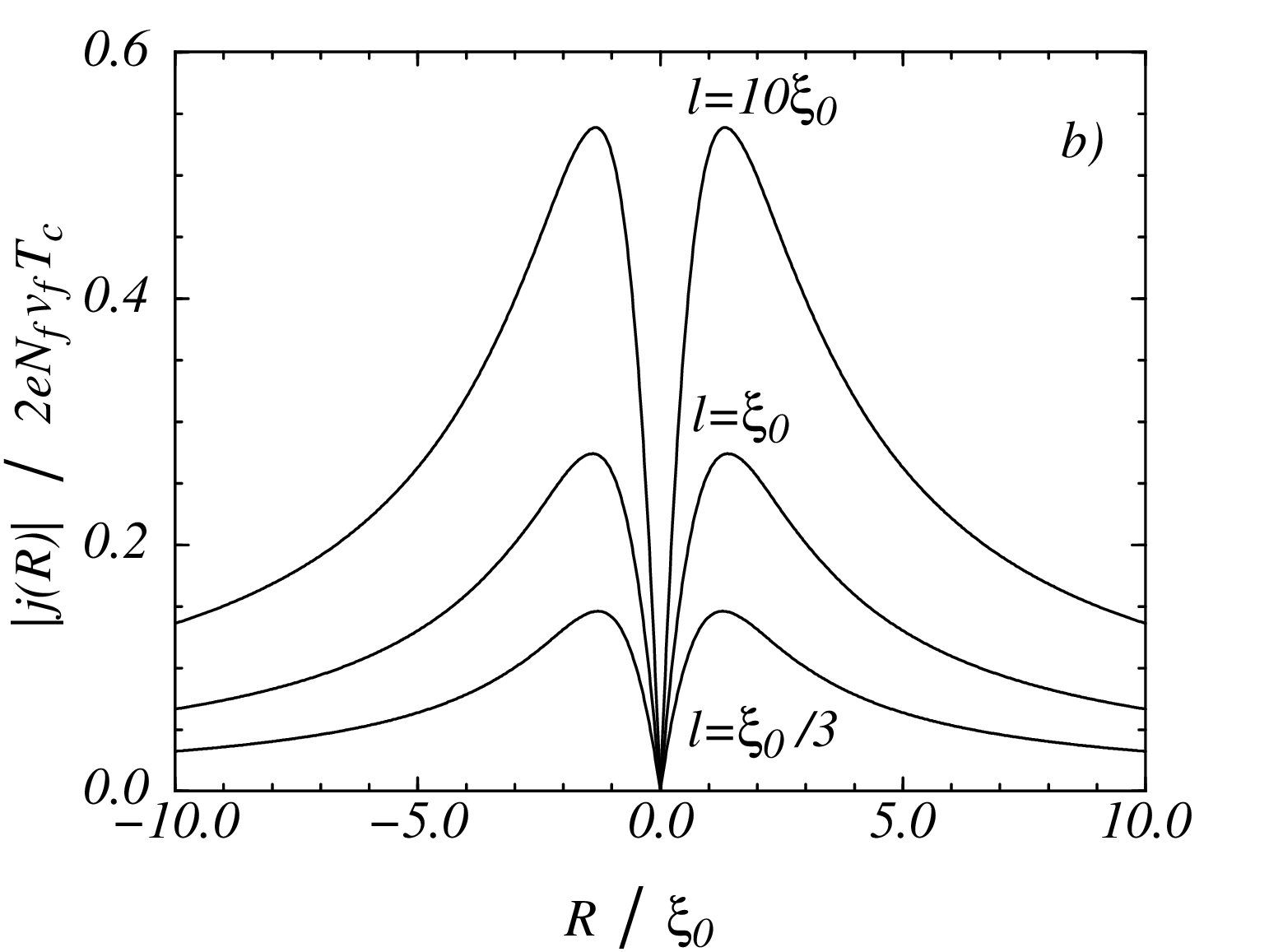}
}
%%%%%%%%%%%%%%%%%%%%%%%%%%%%%%%%%%%%%%%%%%%%%%%%%%%%%%%%
\end{minipage}
\caption[]{ \label{delta} 
Results of  self-consistent calculations of the
modulus of order parameter (a) and the 
current density (b)  at $T=0.3 T_c$
  for  superconductors with  electron mean free path 
  $\ell=10\xi_0$ (clean) to $\ell={1\over 3}\xi_0$ (dirty).
$R$ is the distance from the vortex center measured in units of
the coherence length.}
\end{figure}
Fig.1 shows
the order parameter and the current density  
in the vortex core of an equilibrium vortex 
for different impurity scattering rates $\tau$. 
As expected,
scattering reduces the coherence length and thus the 
size of the core, and has a strong effect on the current
density. Numerical results for the excitation spectrum of
bound and continuum states at the vortex 
together with  
the corresponding  spectral current densities are shown in Fig.2.
The local density of states (per spin) and the spectral current density
are defined by
\begin{eqnarray}\label{dos}
N(\vec{R},\epsilon ) &=& 
N_f \frac{i }{4\pi}\mbox{Tr} \big< \hat\tau_3 \; 
\left( \hat g^R (\vec{p}_f^{\prime},\vec{R}, \epsilon ) 
- \hat g^A (\vec{p}_f^{\prime},\vec{R}, \epsilon ) \right)
 \big> \\
\vec{j}(\vec{R},\epsilon ) &=& 
eN_f \frac{i}{2\pi }  \mbox{Tr}
\big< \hat\tau_3 \vec{v}_f(\vec{p}_f^{\prime})
\left( \hat g^R (\vec{p}_f^{\prime},\vec{R}, \epsilon ) 
- \hat g^A (\vec{p}_f^{\prime},\vec{R}, \epsilon ) \right)
\big>\enspace .
\end{eqnarray}
The zero-energy  bound state is remarkably  broadened 
by impurity scattering. Its width 
 is well approximated by the scattering rate
 $1/(v_f\ell)$,
which is $0.63 T_c$ for $\ell = 10\xi_0$.
The broadening of the bound states decreases with
increasing  energy. The results for the spectral
current density (Fig.2b) show that
nearly all of the
 current
density of the equilibrium vortex 
resides in the energy range of the bound states. 
This  reflects the observation 
that the supercurrents in the vortex core
are predominantly carried by the bound states (Bardeen et al.
1969; Rainer et al. 1996). The physics of the core is
dominated by the bound states, 
in particular also, as we will show,  the  response of the
core to an electric field. 
\begin{figure}[t]
%%%%%%%%%%%%%%%%%%%   F I G U R E   %%%%%%%%%%%%%%%%%%%%
\begin{minipage}[t]{6.0cm}
\centerline{
\epsfxsize5.5cm
\epsffile{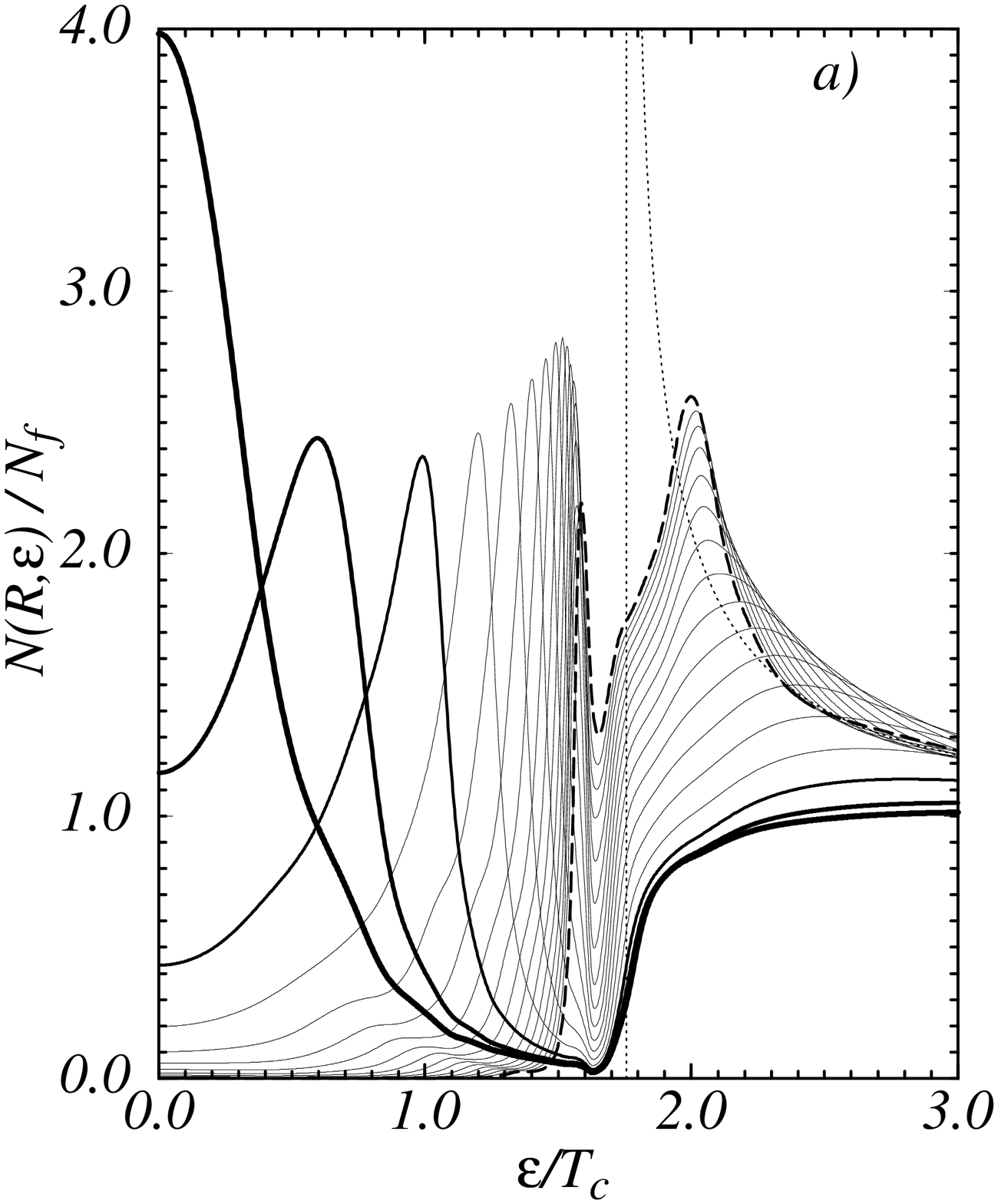}
}
%%%%%%%%%%%%%%%%%%%%%%%%%%%%%%%%%%%%%%%%%%%%%%%%%%%%%%%%
\end{minipage}
\begin{minipage}[t]{6.0cm}
%%%%%%%%%%%%%%%%%%%   F I G U R E   %%%%%%%%%%%%%%%%%%%%
\centerline{
\epsfxsize5.5cm
\epsffile{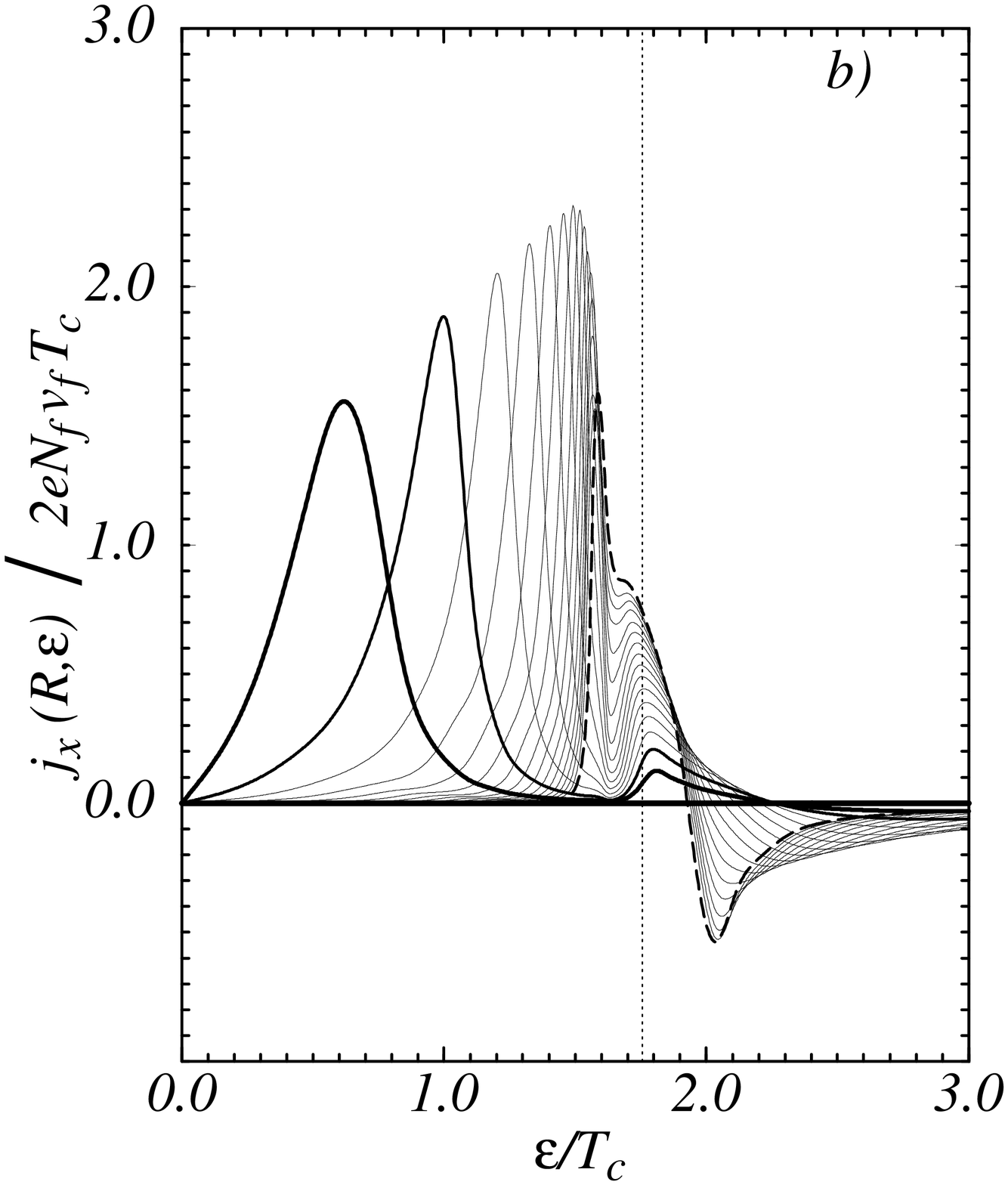}
}
%%%%%%%%%%%%%%%%%%%%%%%%%%%%%%%%%%%%%%%%%%%%%%%%%%%%%%%%
\end{minipage} 
\caption[]{ \label{Nofepss1} 
Local density of states $N( \vec{R},\epsilon )$ (a) and
local spectral current density $j_x( \vec{R},\epsilon)$ (b)
in  the vortex core of a superconductor with
$\ell = 10\xi_0$ at $T=0.3 T_c$. Results are shown for a series
of spatial points on the y-axis, at 
distances 0, $.25\pi\xi_0,\ .5\pi\xi_0,\ \ldots\ , 4\pi\xi_0$ 
from the vortex center.
The thickest full line corresponds to the vortex center, and 
decreasing thickness indicates increasing distance from the
center.  Results for the outermost point ($4\pi\xi_0$) 
are shown as dashed lines.
The thin  dotted lines show the  density of states of the
homogeneous superconductor (a) and
the value of the  bulk gap (b) respectively. 
} 
\end{figure}

\section{Distribution Functions}

The set of  formulas for calculating the quasiclassical 
linear response of a
superconductor 
is  given in a compact notation in 
(\ref{linresp})-(\ref{localneutral}). In this section we transform
these formulas into a more  suitable form for analytical and
numerical calculations. 
The central differential equation of the
quasiclassical response theory is 
the transport equation  (\ref{transpqcl1}). It  
comprises 
12  differential equations for the  components of
the three
 $2\times 2$-Nambu matrices, $\delta\hat g^{R,A,K}$. The number
of differential equations can be reduced significantly 
by using general symmetry
relations (see \cite{Rainer97})  and the
normalization conditions of the quasiclassical theory.
We focus
 on the simplifications which follow  from the
normalization equations 
(\ref{normalize0}) and  (\ref{normalize1}). 
We use the projection operators introduced by  
 Shelankov (1984), 
\begin{equation}\label{project}
\hat{P}^{R,A}_+={1\over 2}\left(\hat 1+{1\over
-i\pi}\hat{g}^{R,A}_0\right)\,, \enspace 
\hat{P}^{R,A}_-={1\over 2}\left(\hat 1-{1\over
-i\pi}\hat{g}^{R,A}_0\right)\enspace . 
\end{equation}
Obviously,
$\hat{P}^{R,A}_{+}+\hat{P}^{R,A}_{-}=\hat{1}$. 
The algebra of the projection operators follows from the
normalization conditions.
\begin{eqnarray}\label{projalgebra}
&&(\hat{P}^{R,A}_{+})^2 =\hat{P}^{R,A}_{+},\ 
(\hat{P}^{R,A}_{-})^2=\hat{P}^{R,A}_{-},\nonumber \\
&&\hat{P}^{R,A}_{+}\hat{P}^{R,A}_{-}=
\hat{P}^{R,A}_{-}\hat{P}^{R,A}_{+}=0 . 
\end{eqnarray}
A key result is
that  the Nambu matrices $\delta\hat{g}^{R,A,K}$ can be
expressed, with the help of Shelankov's projectors,  in
terms of  6 scalar {\em distribution functions}, 
$\delta\gamma^{R,A}$, $\delta\tilde{\gamma}^{R,A}$,
$\delta x^a$ and $\delta\tilde{x}^a$, 
each of which is a  function of $\vec{p}_f$, $\vec{R}$, $\epsilon$,
 $t$, and satisfies a scalar transport equation.
The distribution functions are defined by 
\begin{eqnarray}\label{eschrigRA}
&&\delta\hat{g}^{R,A}\nonumber\\
&&=\mp 2\pi i\left[
 \hat{P}^{R,A}_+\!\otimes\!
\left(
\begin{array}{cc}
 0&\delta\gamma^{R,A}\\
0&0
\end{array}
\right)\!\otimes\! \hat{P}_-^{R,A}-\hat{P}^{R,A}_-\!\otimes\!
\left(
\begin{array}{cc}
 0&0\\
- \delta\tilde\gamma^{R,A}&0
\end{array}
\right)\!\otimes\! \hat{P}_+^{R,A}\right]
 ,
\end{eqnarray}
and
\begin{eqnarray}\label{eschriga}
&&{\delta\hat{g}^{a}}\nonumber\\
&&=
 -2\pi i\left[ \hat{P}^{R}_+\otimes
\left(
\begin{array}{cc}
 \delta x^{a}&0\\
0&0
\end{array}
\right)\otimes \hat{P}_-^{A}+\hat{P}^{R}_-\otimes
\left(
\begin{array}{cc}
 0&0\\
0& \delta\tilde x^{a}
\end{array}
\right)\otimes \hat{P}_+^{A}\right]
\enspace .
\end{eqnarray}
where the {\em anomalous response}, $\delta\hat g^a$, is
defined in terms of $\delta\hat{g}^K,\ \delta\hat{g}^R,
\delta\hat{g}^A$ by 
\begin{eqnarray}\label{defganomalous}
\delta\hat{g}^K=\delta\hat{g}^R\otimes
\tanh(\beta\epsilon/2)-\tanh(\beta\epsilon/2)\otimes
\delta\hat{g}^A +\delta\hat{g}^a\enspace ,
\end{eqnarray}
The transport equations for  the various distribution functions 
follow from
(\ref{transpqcl0}) and (\ref{transpqcl1}) and one  finds
(\cite{Eschrig97}), 
\begin{eqnarray}\label{transpqcls1}
&&i\vec{v}_f\cdot\vec{\nabla}\delta\gamma^{R,A}+2\epsilon
\delta\gamma^{R,A} \nonumber\\
&& +(\gamma^{R,A}_0\tilde\Delta^{R,A}-\Sigma^{R,A})
\otimes\delta\gamma^{R,A}
+\delta\gamma^{R,A}\otimes(\tilde\Delta^{R,A}\gamma^{R,A}_0 +
\tilde\Sigma^{R,A}) \nonumber\\
&&= -\gamma^{R,A}_0\otimes\delta\tilde\Delta^{R,A}\otimes\gamma^{R,A}_0
\!\!+\delta\Sigma^{R,A}\otimes\gamma^{R,A}_0\!\!-\gamma^{R,A}_0\otimes
\delta\tilde\Sigma^{R,A} -\delta\Delta^{R,A}\! , 
\end{eqnarray}
\begin{eqnarray}
&&i\vec{v}_f\cdot\vec{\nabla}\delta\tilde\gamma^{R,A}-2\epsilon
\delta\tilde\gamma^{R,A} \nonumber\\ \label{transpqcls2}
&&+(\tilde\gamma^{R,A}_0\Delta^{R,A}-\tilde\Sigma^{R,A})
\otimes\delta\tilde\gamma^{R,A}
+\delta\tilde\gamma^{R,A}\otimes(\Delta^{R,A}\tilde\gamma^{R,A}_0 +
\Sigma^{R,A}) \nonumber\\
&&=
-\tilde\gamma^{R,A}_0\otimes\delta\Delta^{R,A}\otimes\tilde\gamma^{R,A}_0
\!\!+\delta\tilde\Sigma^{R,A}\otimes\tilde\gamma^{R,A}_0-
\tilde\gamma^{R,A}_0\otimes
\delta\Sigma^{R,A}\!\! -\delta\tilde\Delta^{R,A}\! , \
\end{eqnarray}
\begin{eqnarray}
&&i\vec{v}_f\cdot\vec{\nabla}\delta x^{a} +i\partial_t\delta x^a
%\nonumber\\
\!\!+ (\gamma^R_0 \tilde\Delta^R\!\!-\Sigma^R)\otimes
\delta x^a+\delta x^a\otimes (\Delta^A \tilde\gamma^A
+\Sigma^A)\nonumber\\ \label{transpqcls3}
&& =
\gamma^R_0\otimes\delta\tilde\Sigma^a\otimes\tilde\gamma^A_0
-\delta\Delta^a\otimes\tilde\gamma^A_0 -\gamma^R_0\otimes
\delta\tilde\Delta^a -\delta\Sigma^a \ ,
\end{eqnarray}
\begin{eqnarray}
&&i\vec{v}_f\cdot\vec{\nabla}\delta\tilde x^{a} 
-i\partial_t\delta\tilde x^a
%\nonumber\\
\!\!+ (\tilde\gamma^R_0 \Delta^R\!\!-\tilde\Sigma^R)\otimes
\delta\tilde x^a+\delta\tilde x^a\otimes (\tilde\Delta^A \gamma^A_0
+\tilde\Sigma^A)\nonumber\\ \label{transpqcls4}
&& =
\tilde\gamma^R_0\otimes\delta
\Sigma^a\otimes\gamma^A_0
-\delta\tilde\Delta^a\otimes\gamma^A_0 -\tilde\gamma^R_0\otimes
\delta\Delta^a -\delta\tilde\Sigma^a 
\ .
\end{eqnarray}
We have used  the following short-hand notation for the driving terms in
the transport equations, which includes  
 external potentials, perturbations  and self-energies:
\begin{eqnarray}\label{deforder0}
-{e\over c}\vec{v}_f\cdot\vec{A}\hat\tau_3+\hat\Delta_{mf0}+
\hat\sigma_{i0}^{R,A}=
{\left(
\begin{array}{cc}
\Sigma^{R,A}&\Delta^{R,A}\\
-\tilde\Delta^{R,A}& \tilde\Sigma^{R,A}
\end{array}
\right)}
%, \ \hat\sigma_{i0}^{a}={\left(
%\begin{array}{cc}
%\Sigma^{a}&\Delta^{a} \\
%-\tilde\Delta^{a}& \tilde\Sigma^{a}
%\end{array}\right)}
\ ,
\end{eqnarray}
\begin{eqnarray}\label{deforder1}
\delta\hat\Delta_{mf}+
\delta\hat\sigma_{i}^{R,A}+\delta\hat v=
{\left(
\begin{array}{cc}
\delta\Sigma^{R,A}&\delta\Delta^{R,A}\\
-\delta\tilde\Delta^{R,A}& \delta\tilde\Sigma^{R,A}
\end{array}
\right)}, \ \delta\hat\sigma_{i}^{a}={\left(
\begin{array}{cc}
\delta\Sigma^{a}&\delta\Delta^{a}\\
\delta\tilde\Delta^{a}& -\delta\tilde\Sigma^{a}
\end{array}\right)}\ .
\end{eqnarray}
The functions $\gamma^{R,A}_0$ and $\tilde\gamma^{R,A}_0$ in 
(\ref{transpqcls1})-(\ref{transpqcls4}) are defined by
the following convenient parameterization of the 
equilibrium propagators (\cite{nagat93,schop95}): 
\begin{eqnarray}\label{definegamma}
\hat g^{R,A}_0 = \mp i\pi{1\over 1+\gamma^{R,A}_0\tilde\gamma^{R,A}_0}
\left(
\begin{array}{cc}
1-\gamma^{R,A}_0\tilde\gamma^{R,A}_0&2\gamma^{R,A}_0\\
2\tilde\gamma^{R,A}_0& -(1-\gamma^{R,A}_0\tilde\gamma^{R,A}_0)
\end{array}
\right)\enspace .
\end{eqnarray}
After elimination of the time-dependence in 
(\ref{transpqcls1})-(\ref{transpqcls4}) by 
Fourier transform one is left with four sets of 
ordinary differential
equations along straight trajectories in $R$-space. 
For given right-hand sides these equations are decoupled, and
determine the
distribution functions $\delta\gamma^{R,A},\
\delta\tilde\gamma^{R,A}, \ \delta x^a$ and $\delta\tilde x^a$.
On the other hand, the self-consistency conditions relate the
right-hand sides of (\ref{transpqcls1})-(\ref{transpqcls4}) 
to the solutions of (\ref{transpqcls1})-(\ref{transpqcls4}).
Hence, equations (\ref{transpqcls1})-(\ref{transpqcls4}) may be 
considered either as a large system of linear differential
equations of size six times the chosen number of trajectories,
or as a self-consistency problem. We solved the self-consistency
problem numerically using special algorithms for updating the
right hand sides.
Details of our  numerical schemes 
for a self-consistent  determination of the response functions
are given elsewhere
(\cite{Eschrig97}). 

\section{Response to an A.C. Electric Field}

\begin{figure}[t]
%%%%%%%%%%%%%%%%%%%   F I G U R E   %%%%%%%%%%%%%%%%%%%%
\begin{minipage}[t]{4.0cm}
\centerline{
\epsfxsize3.8cm
\epsffile{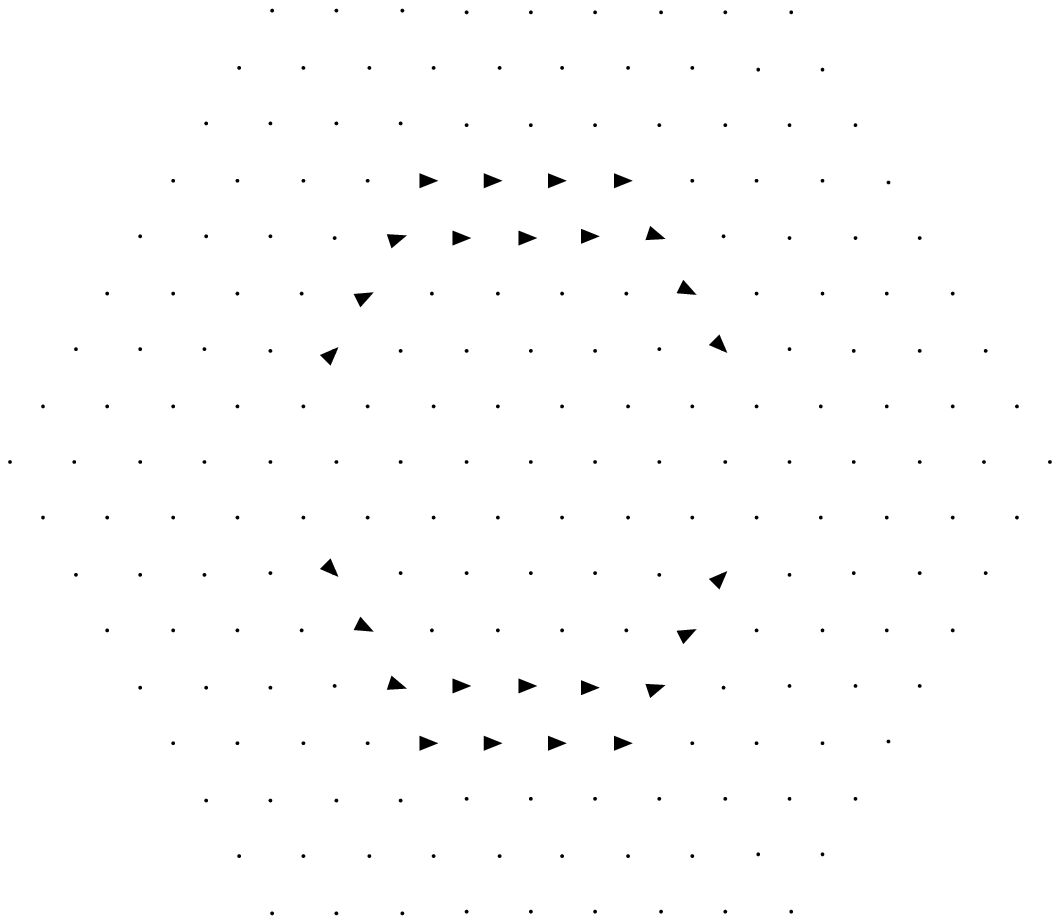}
}
\end{minipage}
\begin{minipage}[t]{4.0cm}
\centerline{
\epsfxsize3.8cm
\epsffile{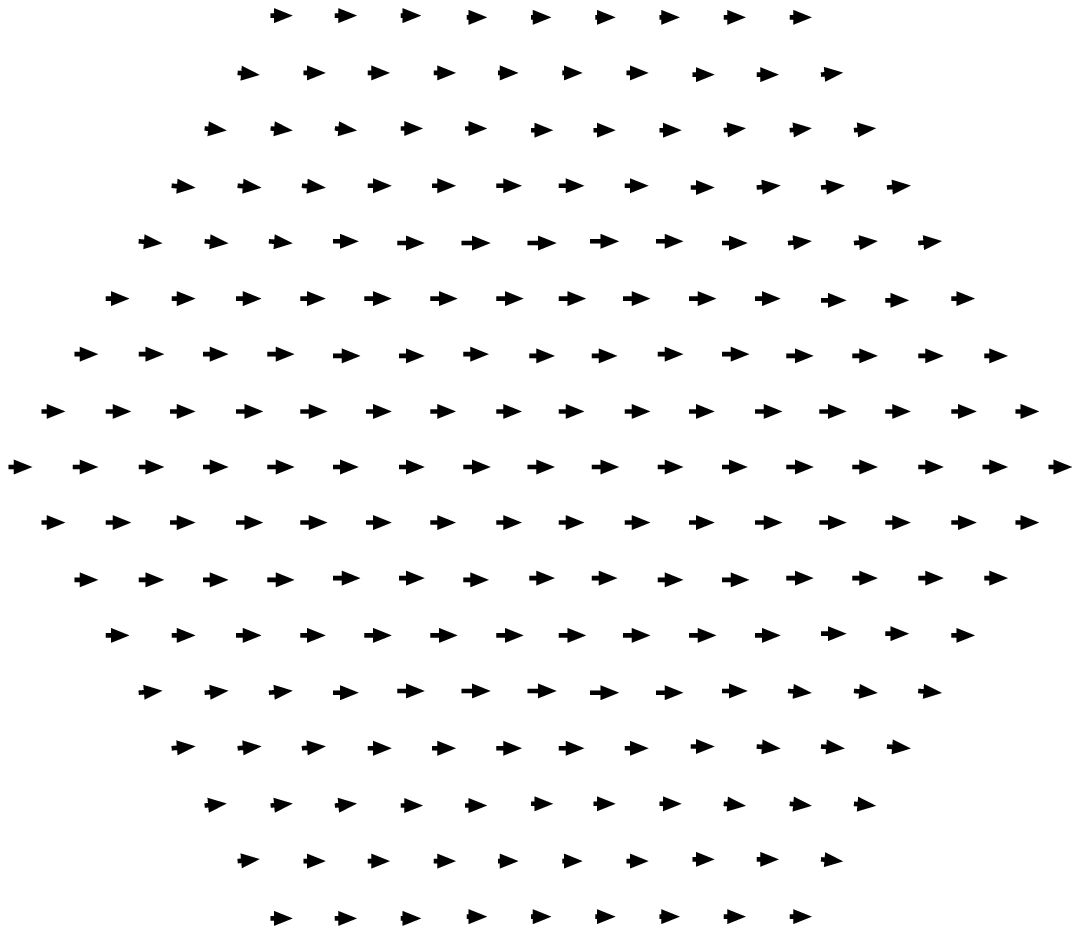}
}
\end{minipage}
\begin{minipage}[t]{4.0cm}
\centerline{
\epsfxsize3.8cm
\epsffile{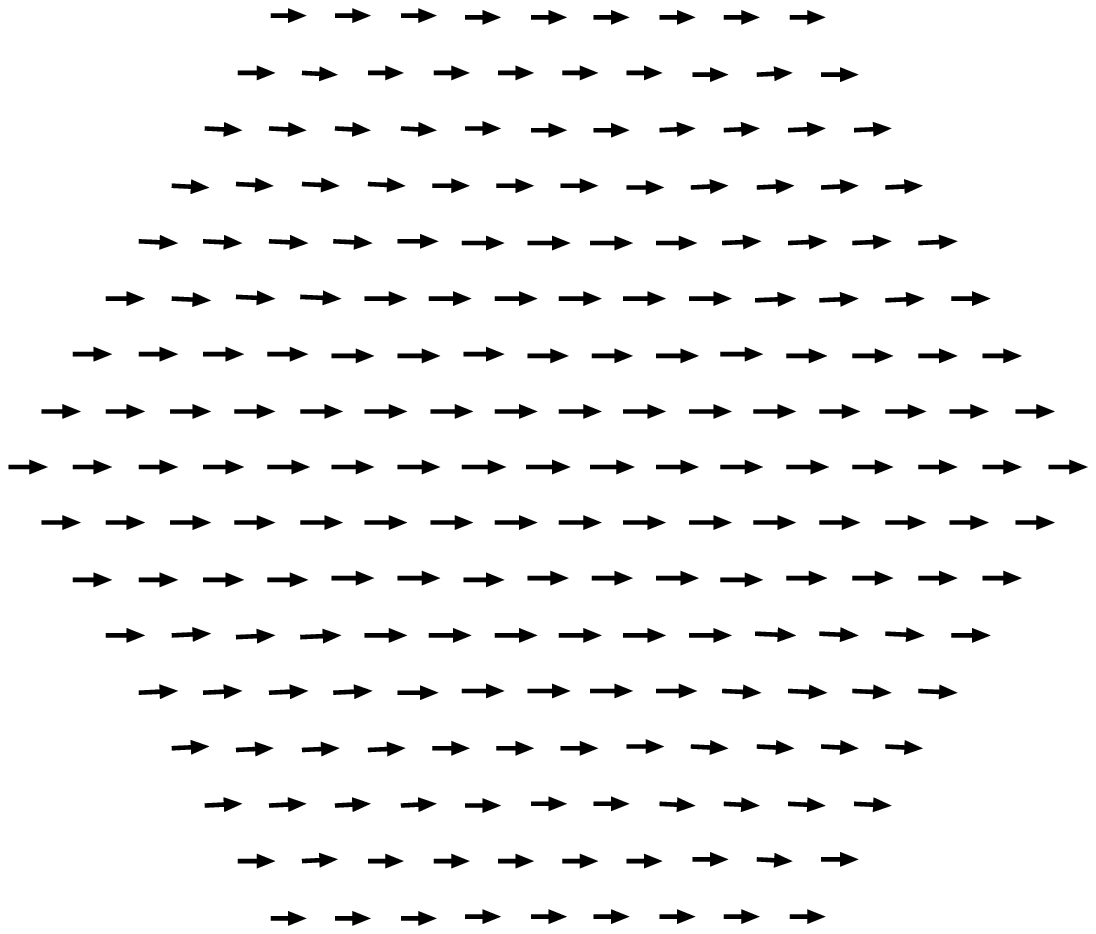}
}
\end{minipage}\\[3mm]
\begin{minipage}[t]{4.0cm}
\centerline{
\epsfxsize3.8cm
\epsffile{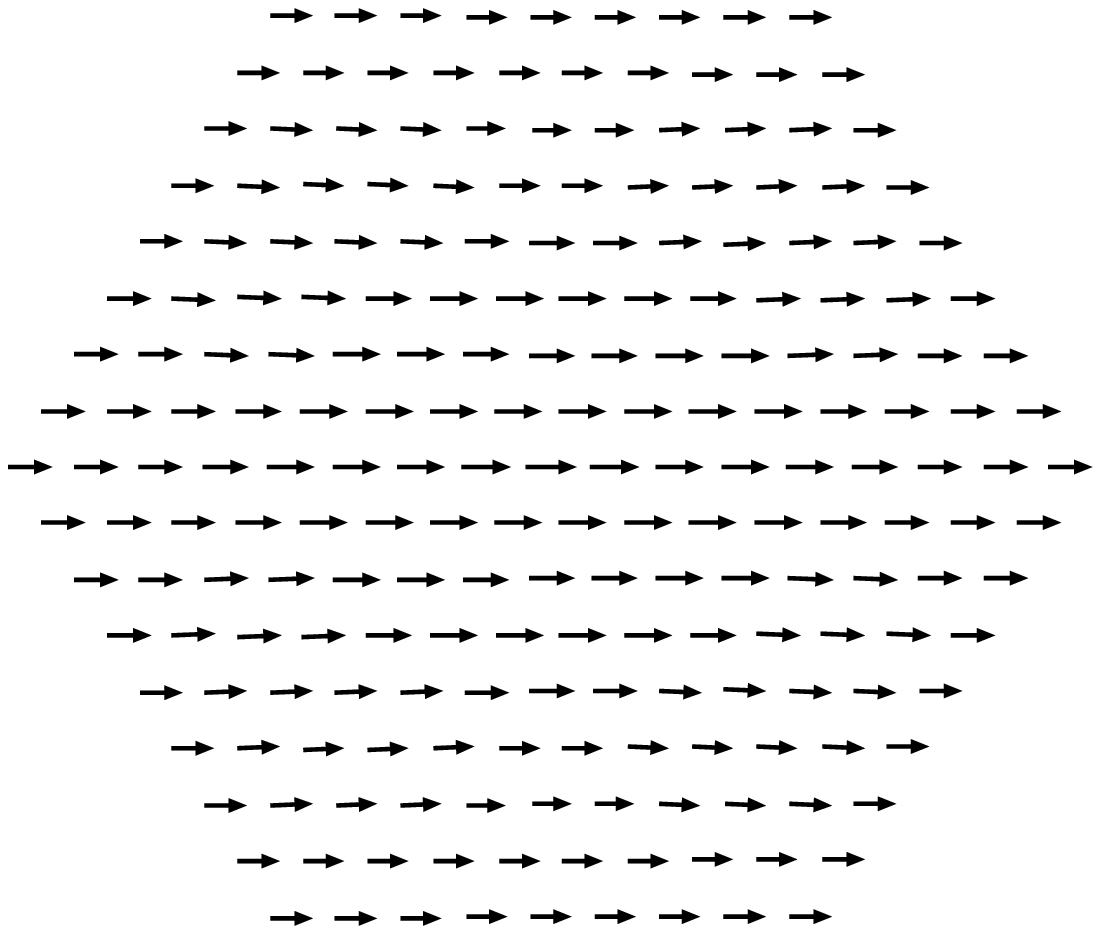}
}
\end{minipage}
\begin{minipage}[t]{4.0cm}
\centerline{
\epsfxsize3.8cm
\epsffile{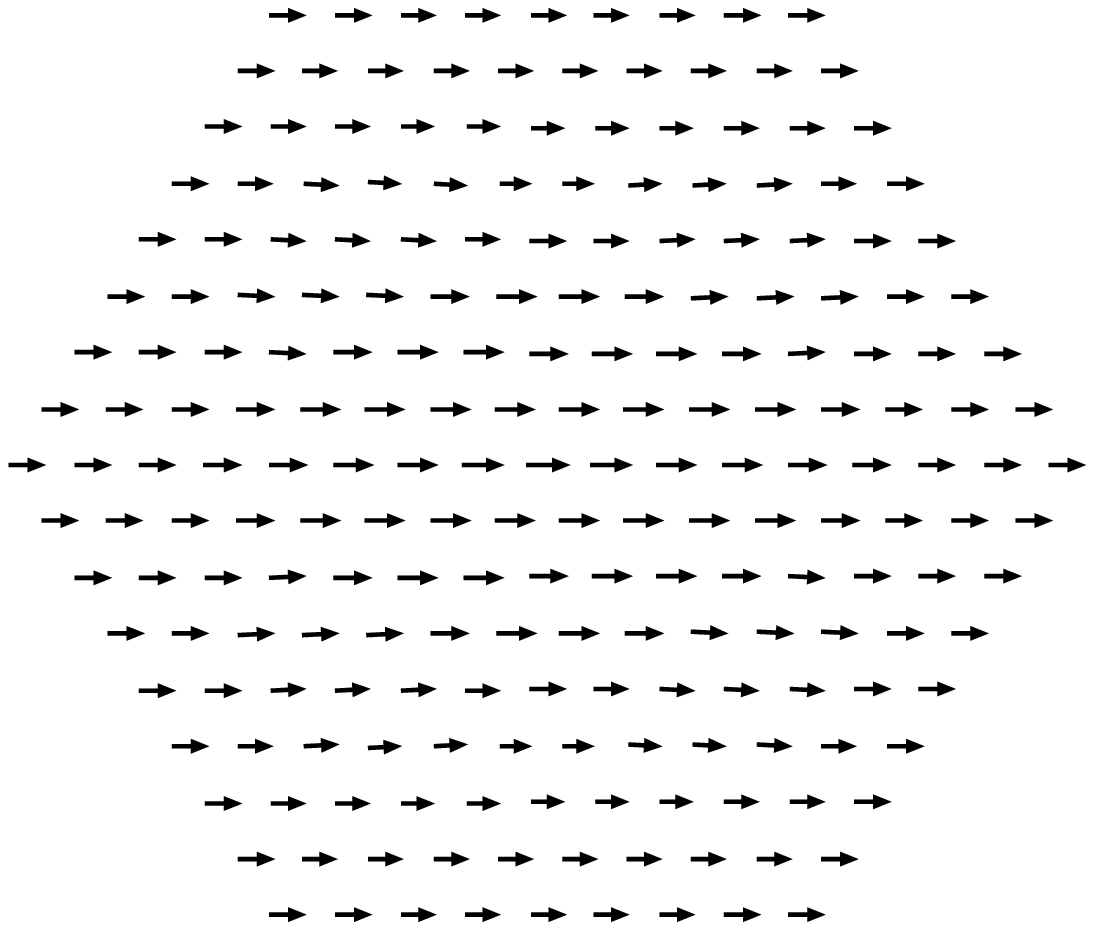}
}
\end{minipage}
\begin{minipage}[t]{4.0cm}
\centerline{
\epsfxsize3.8cm
\epsffile{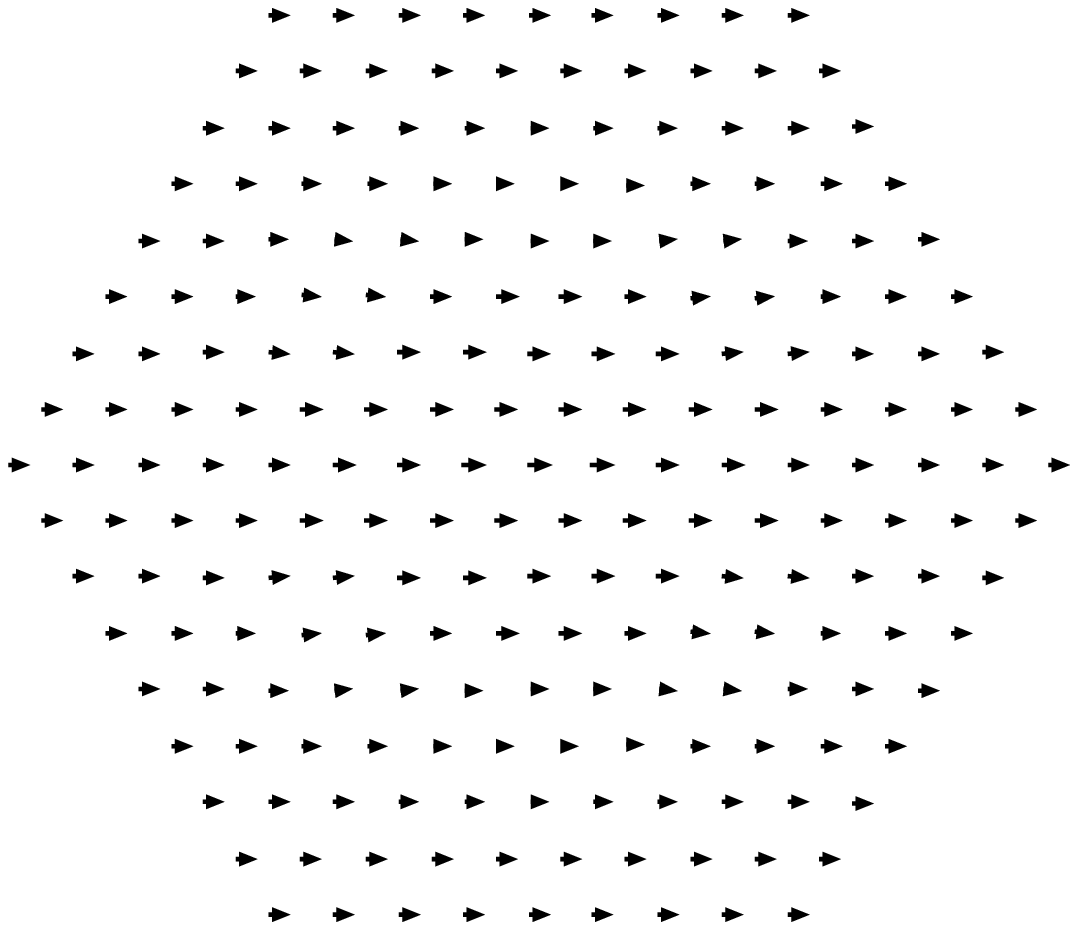}
}
\end{minipage}
\caption[]{ \label{patt2} 
Snapshots of the time-dependent current pattern 
in the core of a pancake vortex at successive times,  
$t =0, {1\over 12}{\cal T},{2\over 12}{\cal T}, 
{3\over 12}{\cal T}, {4\over 12}{\cal T}, {5\over 12}{\cal T}$ 
(upper left pattern to lower right pattern) . 
The  arrows  show the
current density induced by 
 an {\it a.c.} electric field in $x$-direction,
${E_x}(t)=\delta{E} \cos(\omega t)$,
of frequency $\omega={2\pi/ {\cal T}}=1.5\Delta$.
The data are calculated in
linear response approximation 
for a clean superconductor ($\ell = 10 \xi_0 $)
at $T=0.3T_c$.
Current patterns in the second half-period,  
 $t={1\over 2}{\cal T}-{11\over 12}{\cal T}$,  are obtained 
from the patterns in the first half-period 
by reversing the directions of the currents.
The distance between two neighboring points on the hexagonal grid 
is $0.25 \pi \xi_0$.
}
%%%%%%%%%%%%%%%%%%%%%%%%%%%%%%%%%%%%%%%%%%%%%%%%%%%%%%%%
\end{figure}

\begin{figure}[t]
%%%%%%%%%%%%%%%%%%%   F I G U R E   %%%%%%%%%%%%%%%%%%%%
\begin{minipage}[t]{4.0cm}
\centerline{
\epsfxsize3.8cm
\epsffile{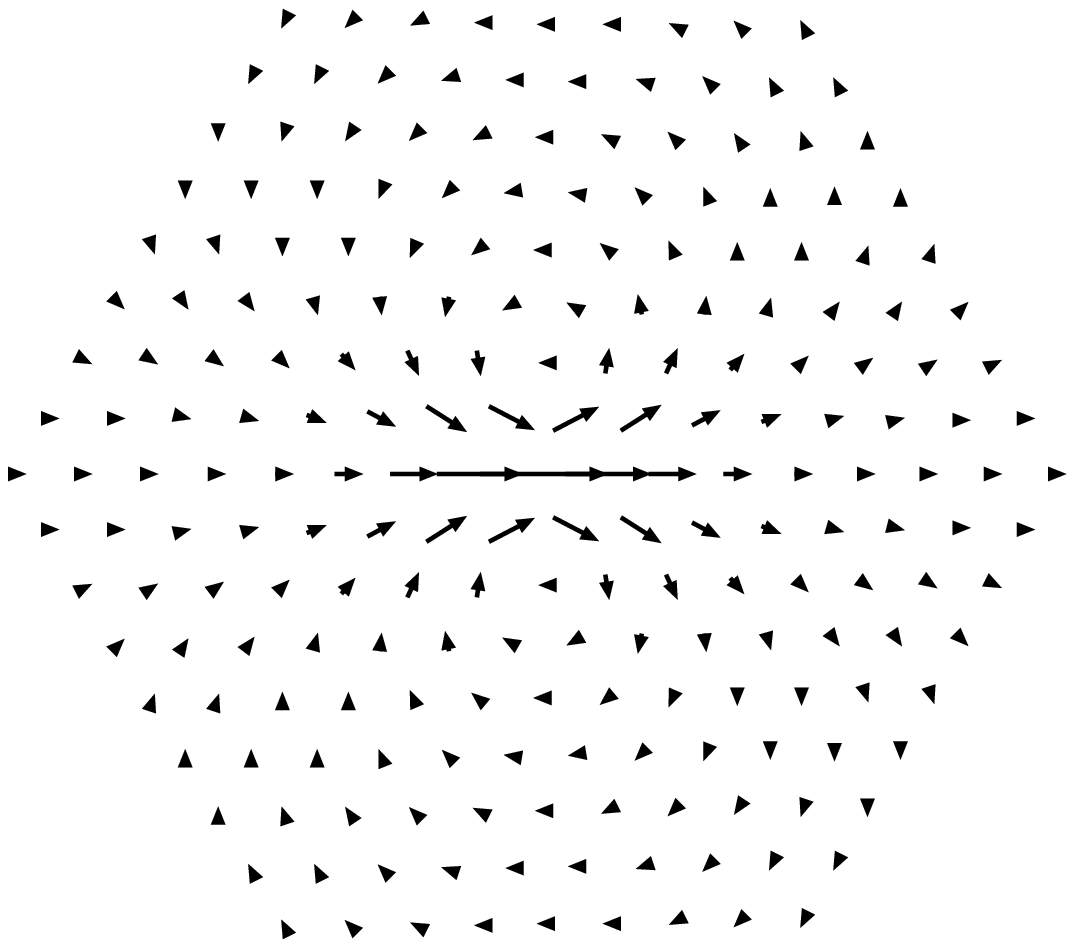}
}
\end{minipage}
\begin{minipage}[t]{4.0cm}
\centerline{
\epsfxsize3.8cm
\epsffile{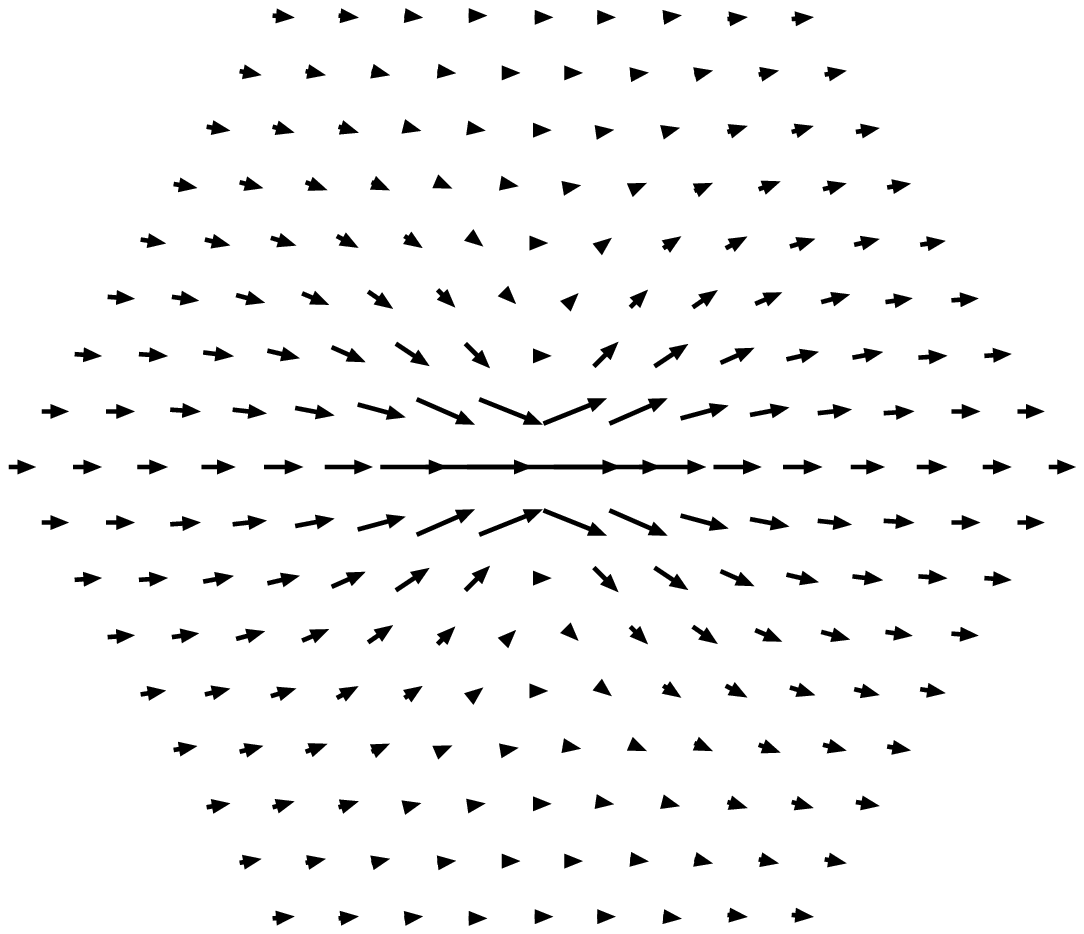}
}
\end{minipage}
\begin{minipage}[t]{4.0cm}
\centerline{
\epsfxsize3.8cm
\epsffile{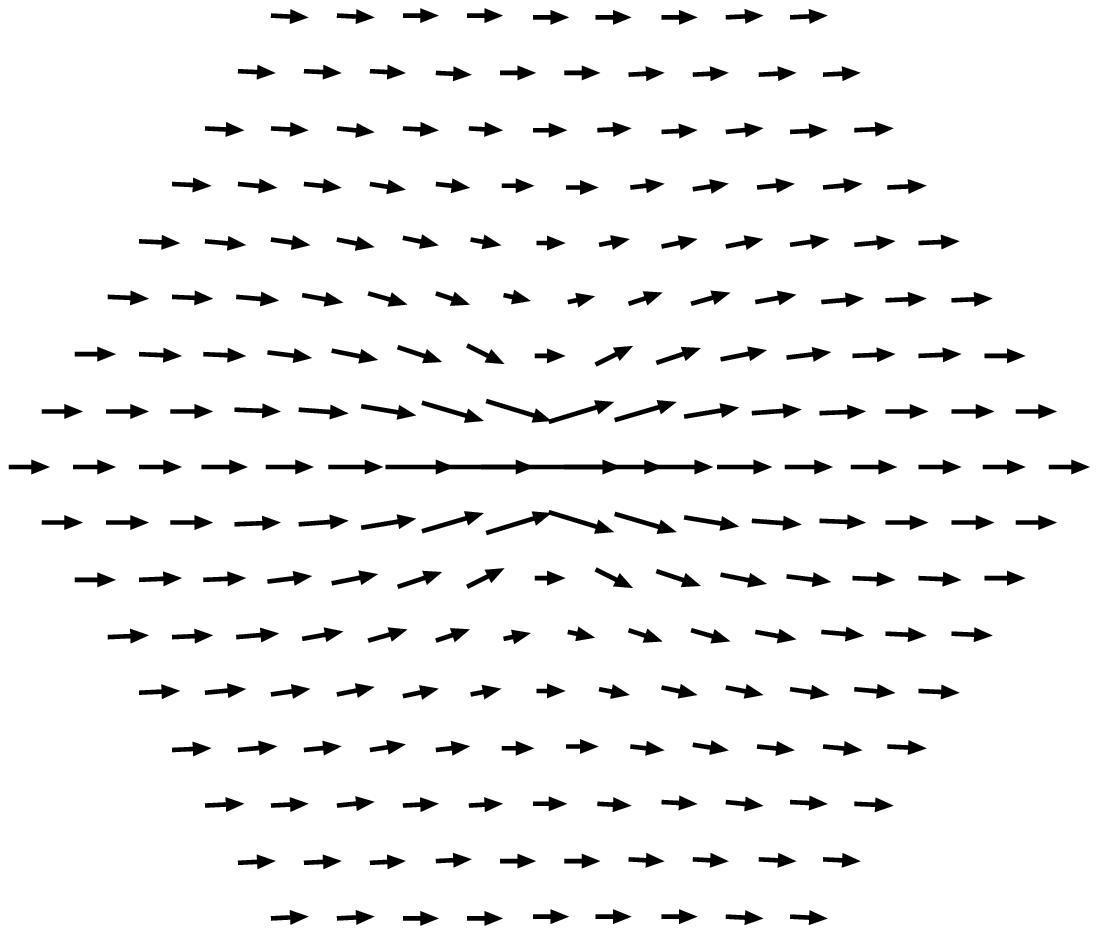}
}
\end{minipage}\\[3mm]
\begin{minipage}[t]{4.0cm}
\centerline{
\epsfxsize3.8cm
\epsffile{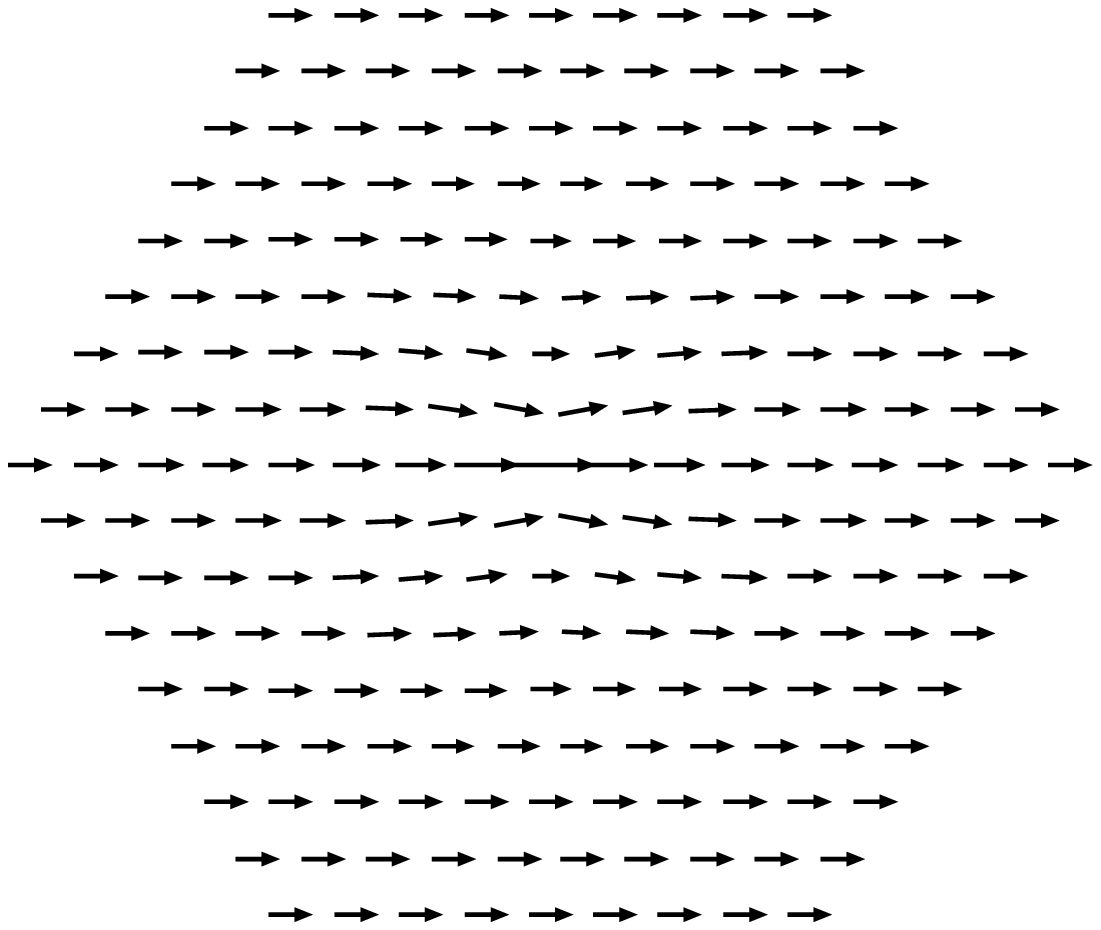}
}
\end{minipage}
\begin{minipage}[t]{4.0cm}
\centerline{
\epsfxsize3.8cm
\epsffile{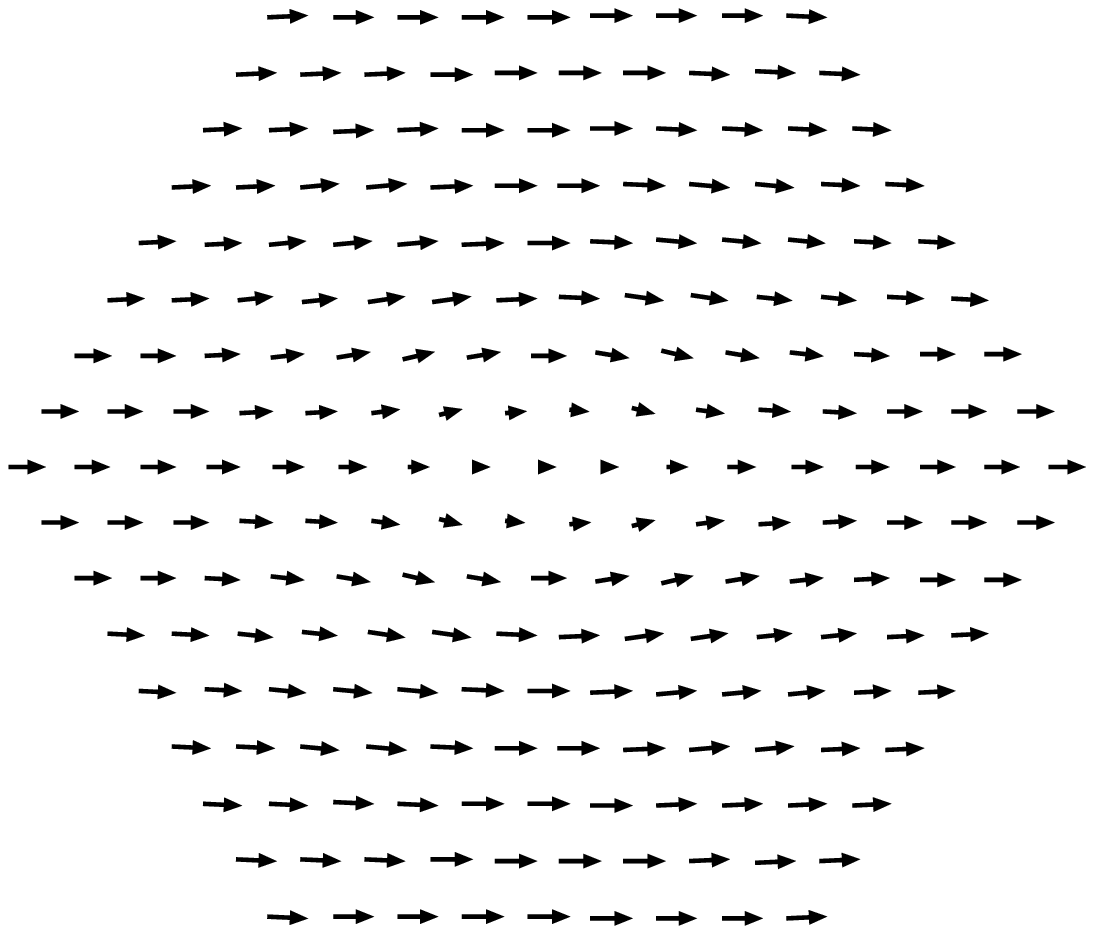}
}
\end{minipage}
\begin{minipage}[t]{4.0cm}
\centerline{
\epsfxsize3.8cm
\epsffile{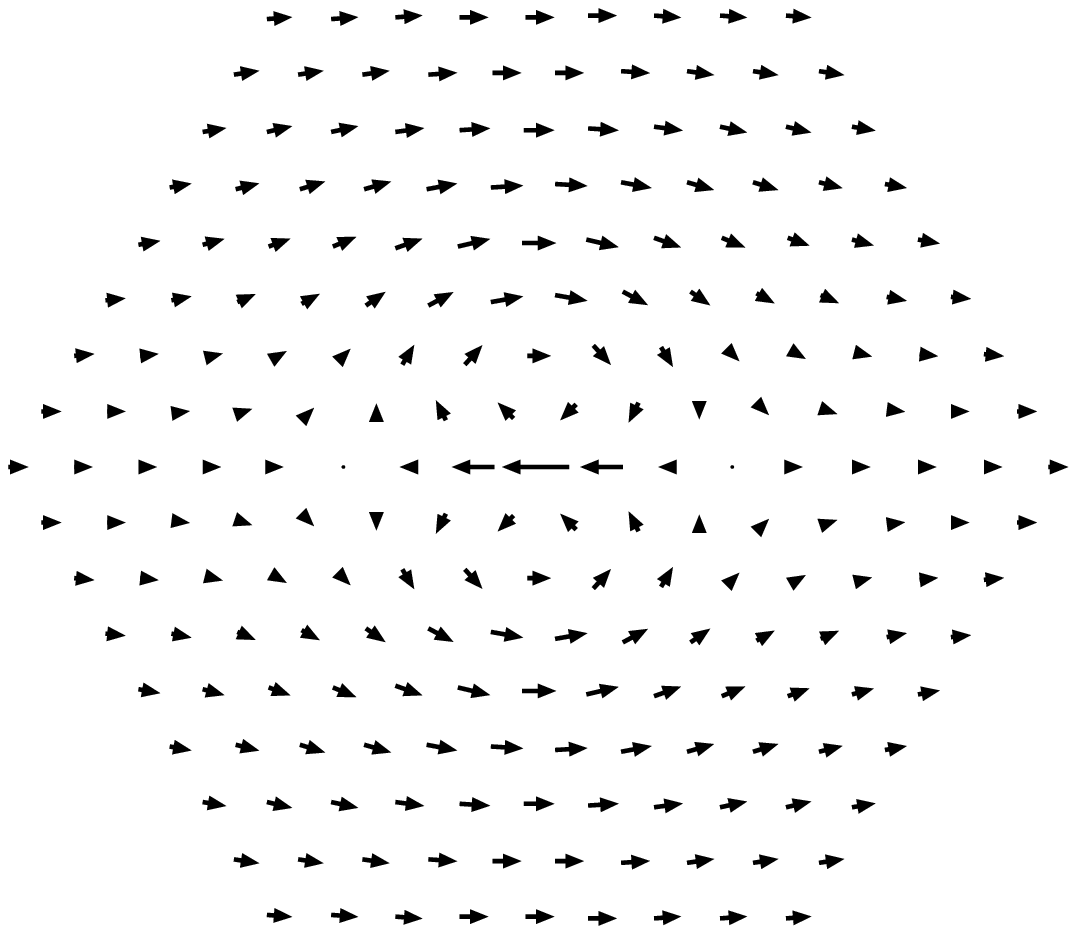}
}
\end{minipage}
\caption[]{ \label{patt1} 
The same as in fig. \ref{patt2}, but
for a smaller external frequency, $\omega = 0.3 \Delta$.
The length of the current vectors is scaled down by a factor 5
compared to those in   fig. \ref{patt2}.
}
%%%%%%%%%%%%%%%%%%%%%%%%%%%%%%%%%%%%%%%%%%%%%%%%%%%%%%%%
\end{figure}

We  consider a pancake vortex in a layered s-wave superconductor
and calculate 
the  response of the electric current density,
$\delta\vec{j}(\vec{R};t)$, to  a small  homogeneous
{\it a.c.} electric field, 
$\delta \vec{E}^{\omega}(t)=\delta\vec{E}\exp(-i\omega t)$.
The results presented in
this section are obtained by solving numerically the 
quasiclassical transport equations 
(\ref{transpqcls1})-(\ref{transpqcls4}) 
together with the self-consistency
equations (\ref{lgapequation1}), (\ref{respborn1}), 
and the condition of local charge neutrality
(\ref{localneutral}). The calculation gives  the local
conductivity tensor, $\sigma_{ij}(\vec{R},\omega)$, defined by
\begin{equation}
\delta j_i(\vec{R},t)=
\sigma_{ij}(\vec{R},\omega)\delta E^{\omega}_j(t)\enspace .
\end{equation}
Figures \ref{patt2} and \ref{patt1}  show the time development of the 
current pattern 
induced by  an oscillating electric field in $x$-direction with
time dependence  
$\delta {E}_x(t)=\delta{E}\cos(\omega t)$. Results are given 
for a medium range frequency ($\omega=1.5\Delta$) and a low frequency
($\omega=0.3\Delta$). 
 Two features should be emphasized.
At medium and higher frequencies the current flow induced by the
electric field is to good 
approximation 
uniform in
space and  phase shifted by $\pi/2$ (non-dissipative currents).
 The phase shift of
$\pi/2$ is the consequence of  a  predominantly 
imaginary conductivity at frequencies above $\Delta$, as
shown in Fig.5. 
The current pattern at
low frequencies (Fig.4) is qualitatively different.
At the vortex center the current
is phase shifted by $\approx \pi/4$ in accordance with 
 the conductivity  
at $\omega=0.3\Delta$, which has about equal real (dissipative)
and imaginary (non-dissipative)
parts, whereas further away from the center the conductivity
becomes more and more 
non-dissipative. Fig.4   shows that  
the current flow at low frequencies is  non-uniform.
The dissipative currents exhibit a dipolar structure
with  enhanced currents  at the
vortex center, and back-flow currents  away from the  center. 
On the other hand, non-dissipative currents are approximately uniform.

\begin{figure}[t]
%%%%%%%%%%%%%%%%%%%   F I G U R E   %%%%%%%%%%%%%%%%%%%%
\begin{minipage}[t]{6.0cm}
\centerline{
\epsfxsize5.5cm
\epsffile{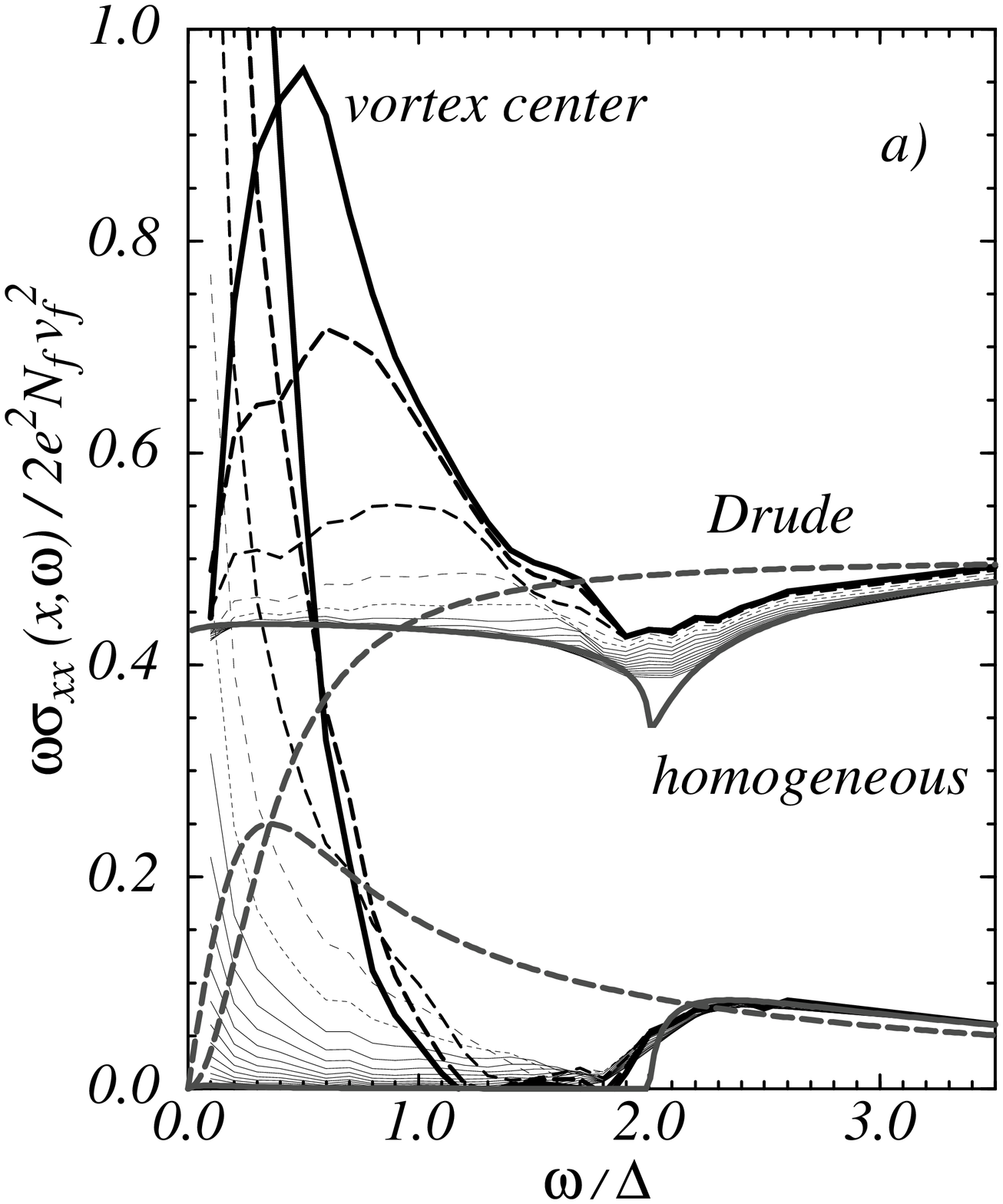}
}
\end{minipage}
\hfill
\begin{minipage}[t]{6.0cm}
\centerline{
\epsfxsize5.5cm
\epsffile{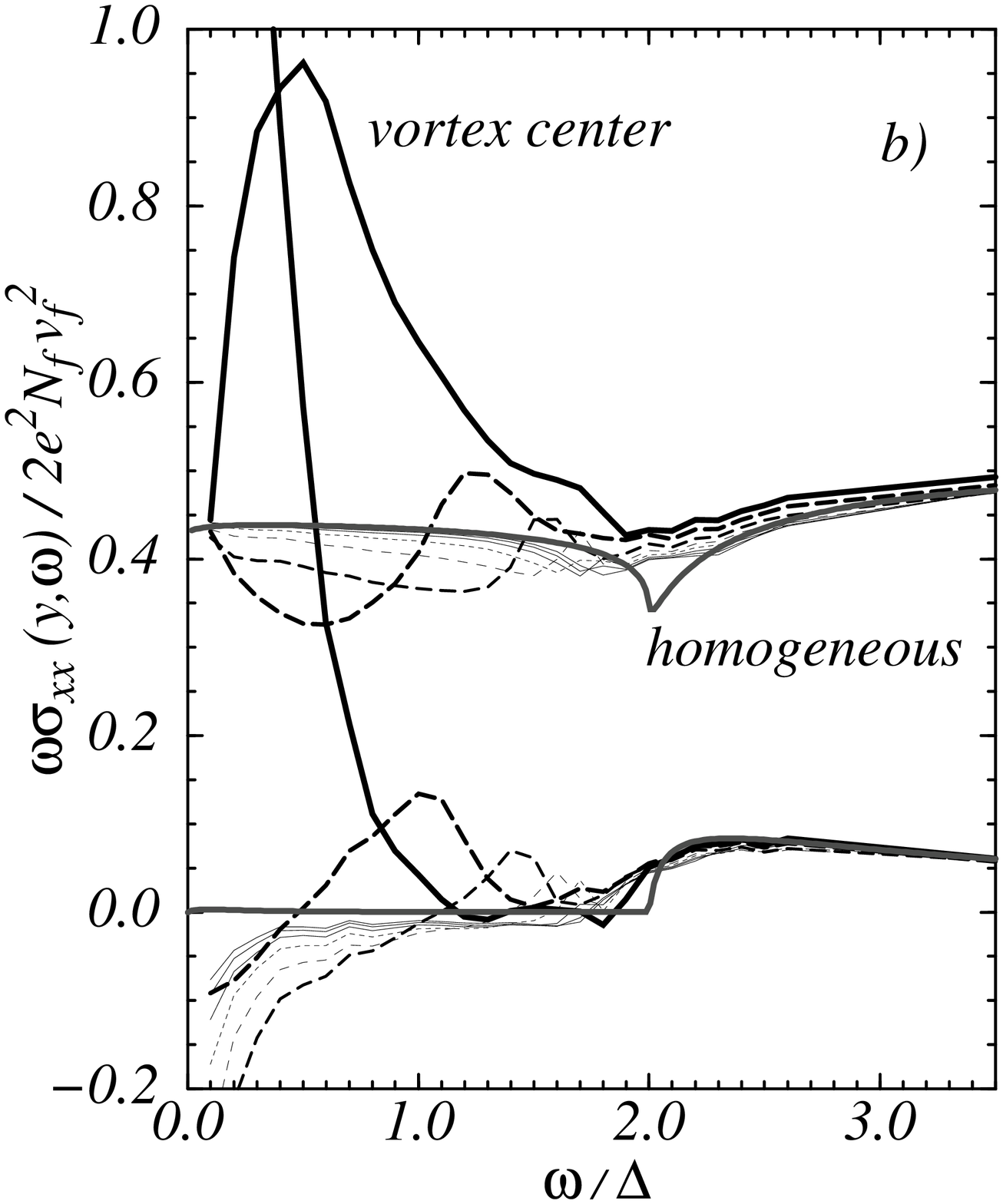}
}
\end{minipage}
\caption[]{ \label{condx} 
Frequency dependence of the real part (`lower' curves) 
and imaginary part (`upper'
curves) of the local conductivity $\sigma_{xx}$
 for a superconductor with
parameters $T=0.3T_c$, $\ell = 10 \xi_0 $. 
For convenience, the conductivities are multiplied by $\omega$.
The full black
curves give  the conductivity at the vortex center, and the
series of dashed lines  with decreasing intensity
the conductivity at  increasing 
distance from the center. 
Fig.\ref{condx}a presents data at  points along the $x$-axis 
(in steps of $(\pi/4)\xi_0$), 
and Fig.\ref{condx}b at points along the $y$-axis
(in steps of $(\sqrt{3}\pi/4) \xi_0$).
The dashed grey lines show the normal state Drude conductivity,
and the full grey lines the conductivity
of the  homogeneous superconductor.
}
%%%%%%%%%%%%%%%%%%%%%%%%%%%%%%%%%%%%%%%%%%%%%%%%%%%%%%%%
\end{figure}
\begin{figure}[t]
%%%%%%%%%%%%%%%%%%%   F I G U R E   %%%%%%%%%%%%%%%%%%%%
\begin{minipage}[t]{4.0cm}
\centerline{
\epsfxsize3.8cm
\epsffile{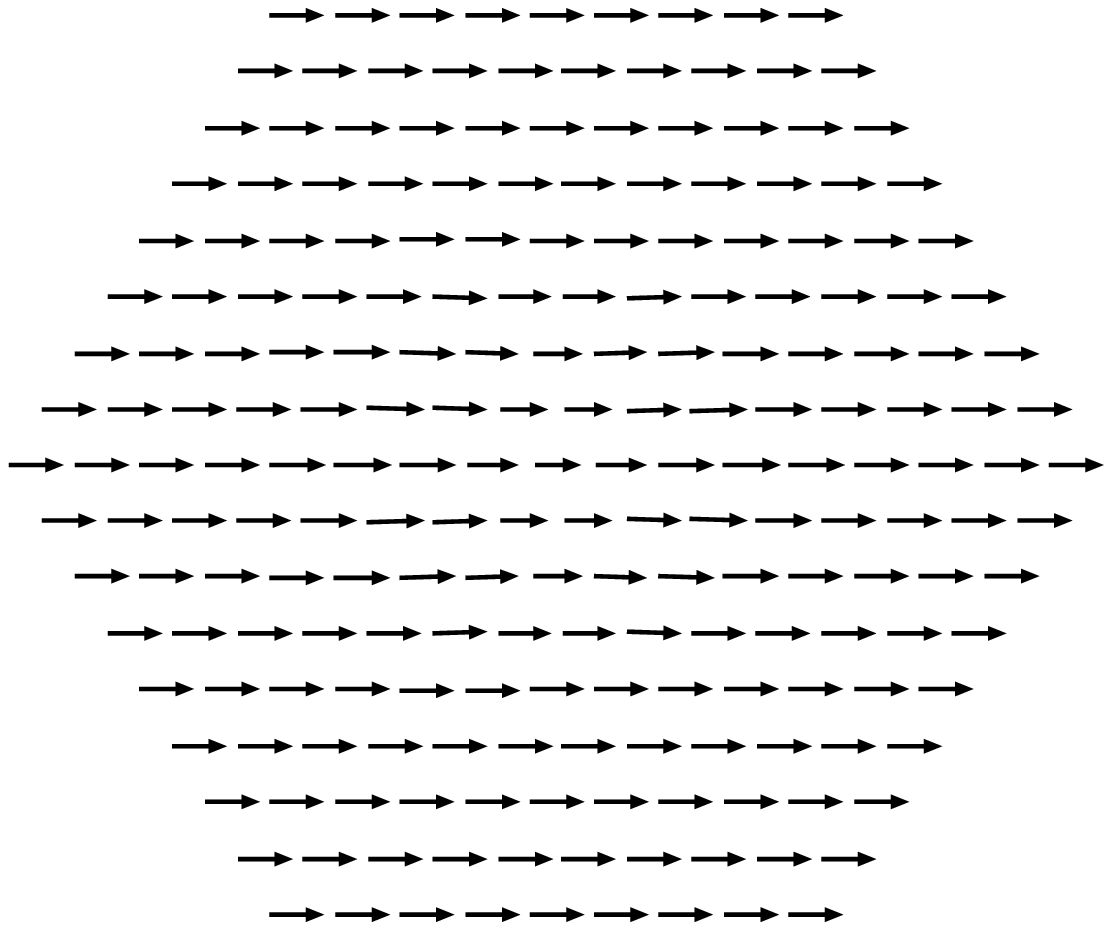}
}
\end{minipage}
\begin{minipage}[t]{4.0cm}
\centerline{
\epsfxsize3.8cm
\epsffile{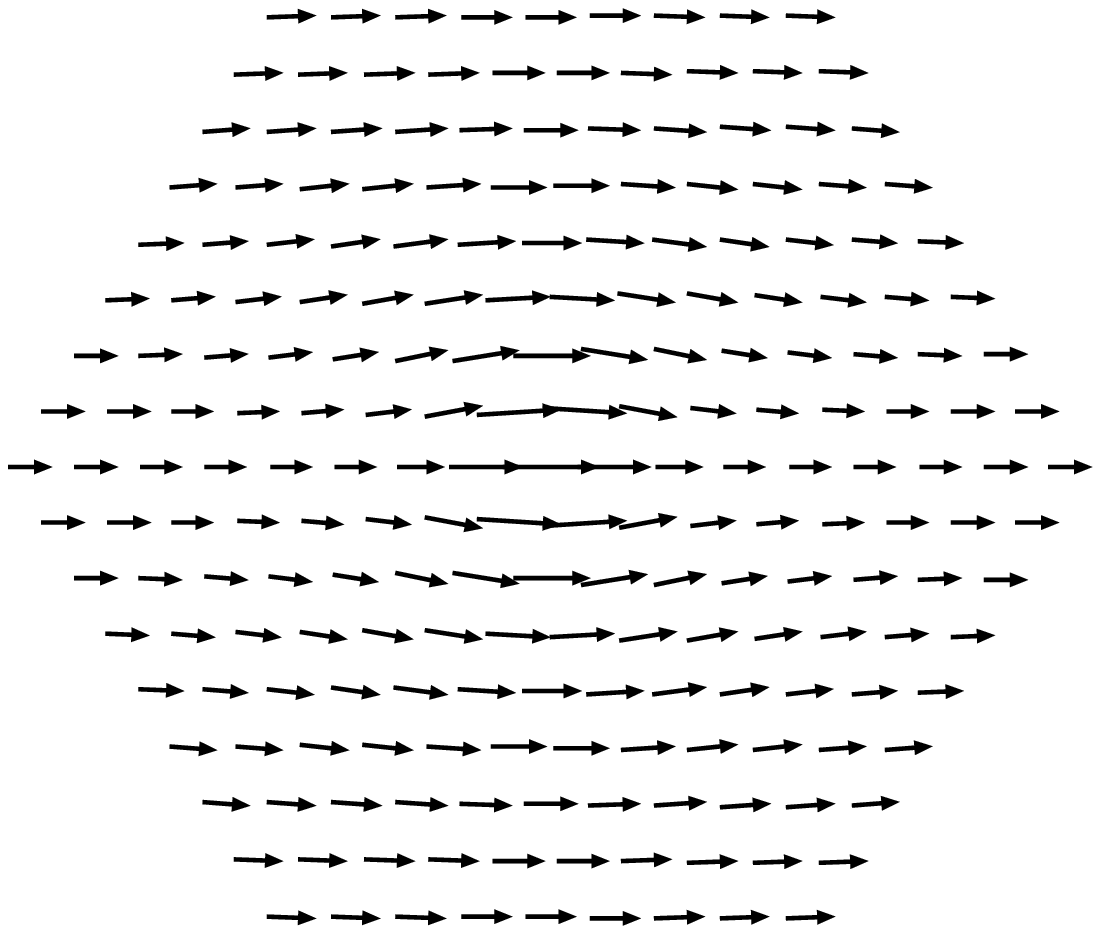}
}
\end{minipage}
\begin{minipage}[t]{4.0cm}
\centerline{
\epsfxsize3.8cm
\epsffile{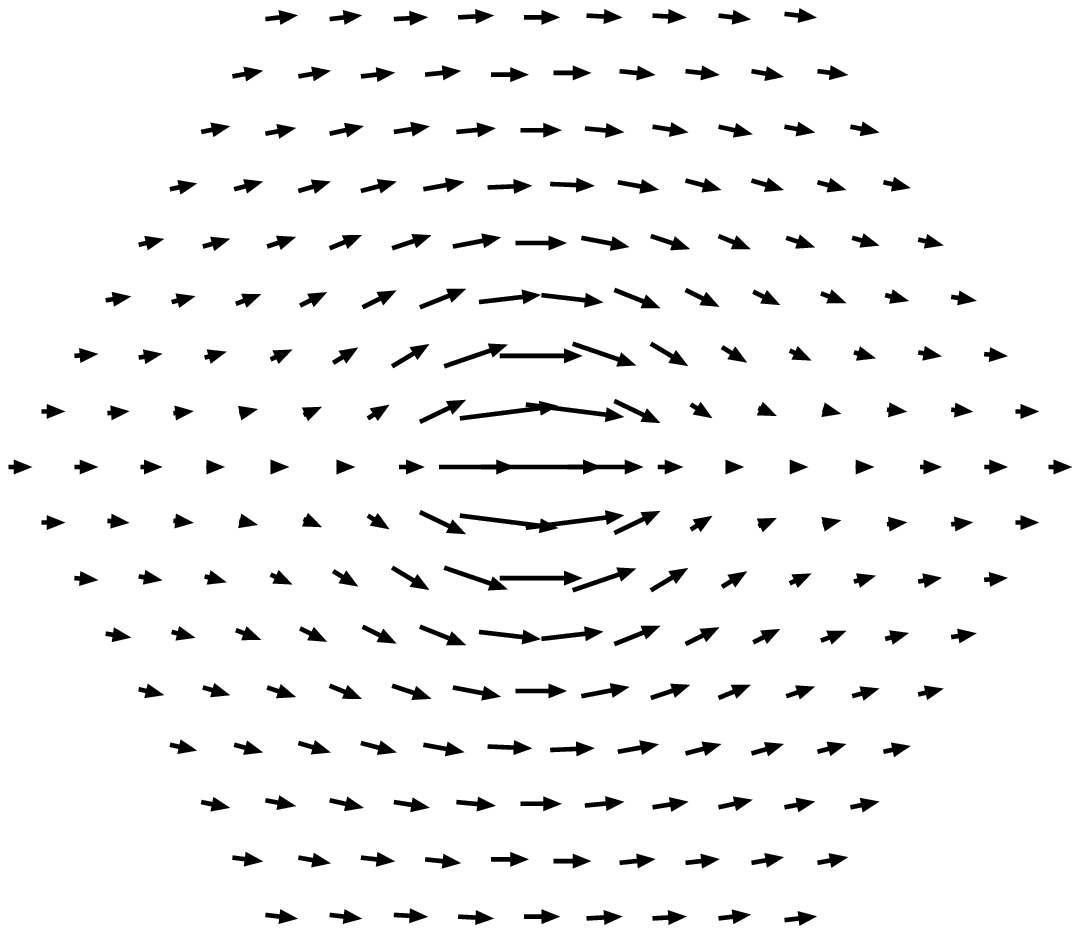}
}
\end{minipage}\\[3mm]
\begin{minipage}[t]{4.0cm}
\centerline{
\epsfxsize3.8cm
\epsffile{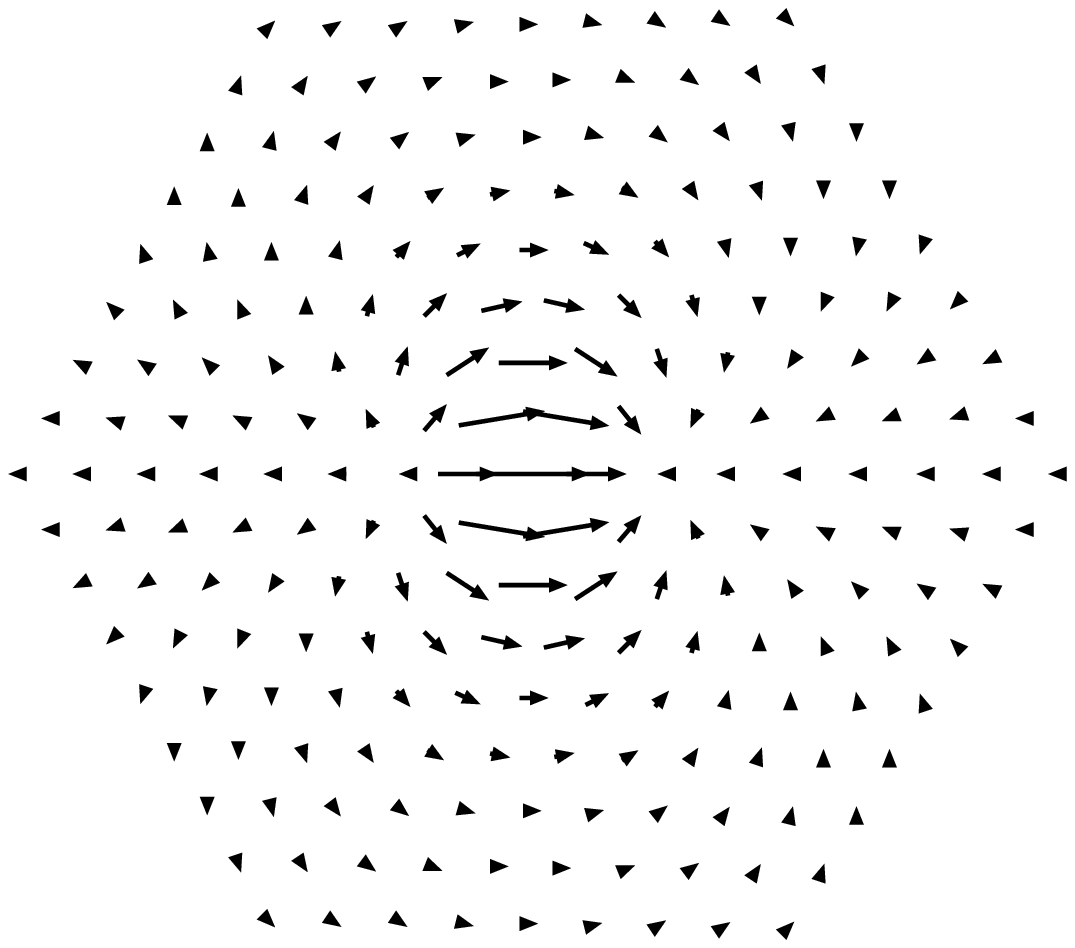}
}
\end{minipage}
\begin{minipage}[t]{4.0cm}
\centerline{
\epsfxsize3.8cm
\epsffile{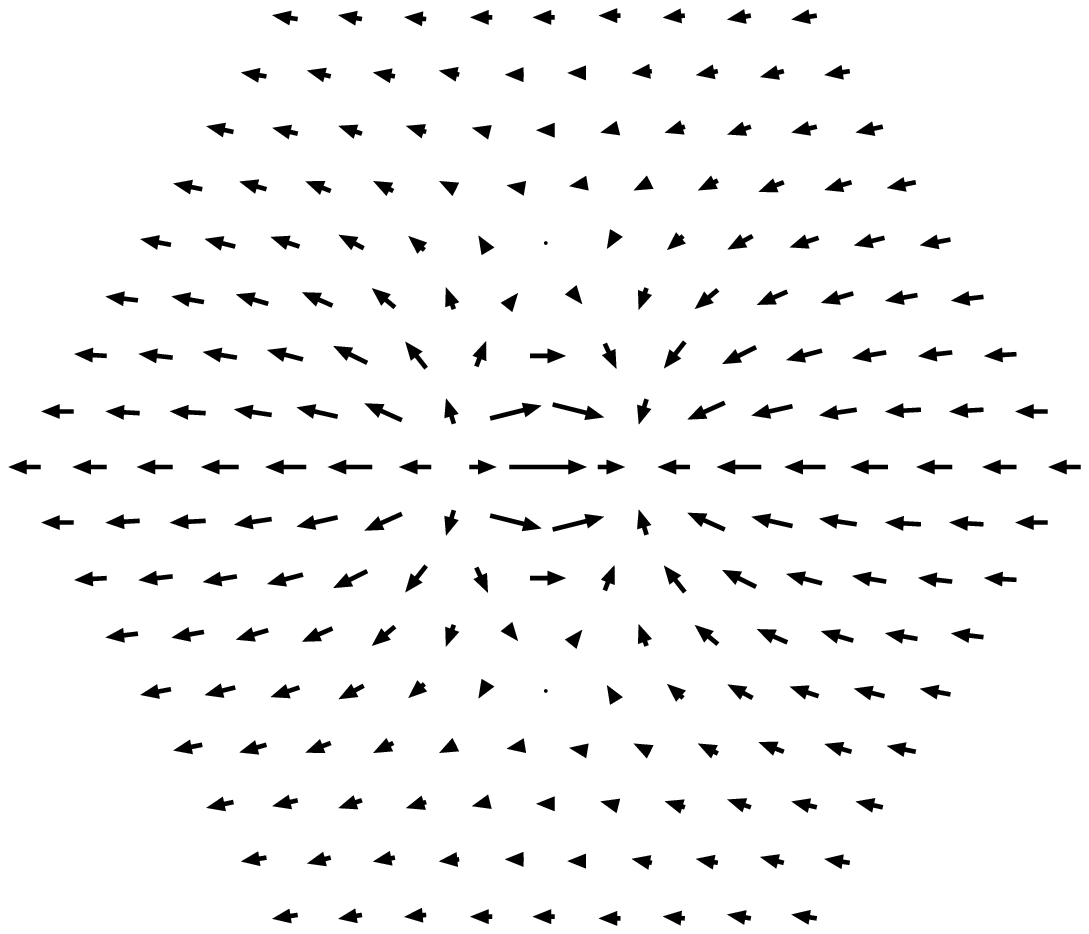}
}
\end{minipage}
\begin{minipage}[t]{4.0cm}
\centerline{
\epsfxsize3.8cm
\epsffile{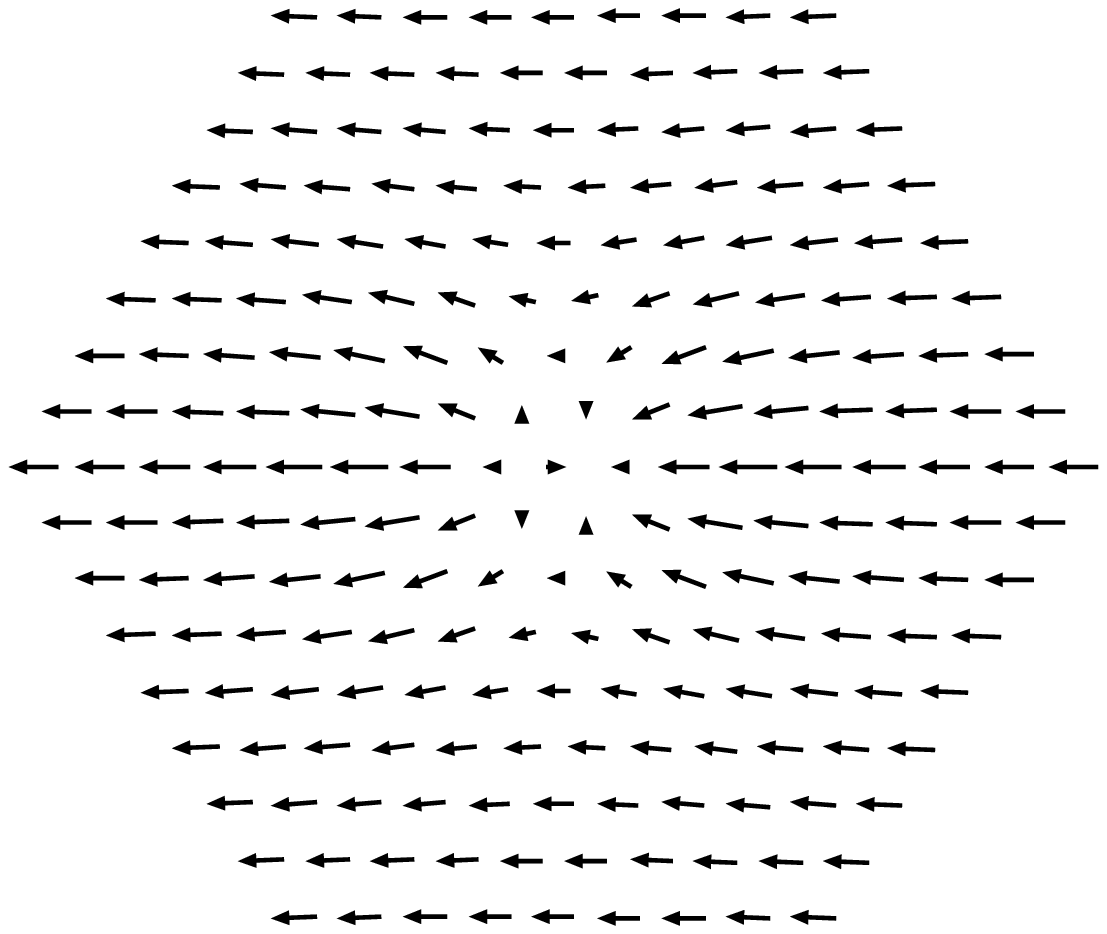}
}
\end{minipage}
\caption[]{ \label{fpatt1} 
Local electrical field for a superconductor with
parameters $T=0.3T_c$, $\ell = 10 \xi_0 $,
and an external frequency of $\omega = 0.3 \Delta (T)$.
Each field pattern is a snapshot at time $t$ varying
from 0 to half of time periode in timesteps of 
$\frac{1}{12}$ periode.
The external electrical field 
$\delta\vec{E}^{\omega}(t)=\delta\vec{E} \cos(\omega t)$
is maximal and points in positive
$x$-direction for the first picture ($t=0$).
In the first picture of the second line it is zero.
The distance between two points in the grid 
corresponds to $0.25 \pi \xi_0$.
}
%%%%%%%%%%%%%%%%%%%%%%%%%%%%%%%%%%%%%%%%%%%%%%%%%%%%%%%%
\end{figure}

The frequency dependence of the local conductivity, 
$\sigma_{xx}(\vec{R},\omega)$,  for $\vec{R}$ along 
the x- and y-axis is shown in figures \ref{condx}a,b.  
These figures
include, for comparison, the Drude conductivity of the normal
state and the conductivity of the homogeneous bulk superconductor. 
A  significant   feature of conductivity in the core 
is the strong increase of ${\cal R}e\, \sigma$ at 
low frequencies. The conductivity is in this
frequency range much  larger than the normal state
Drude conductivity and the exponentially small
conductivity of the bulk s-wave superconductor.
The real part of conductivity scales at low frequencies 
like $1/\omega^2$. 
Its value at
at the vortex center  is 
 $69.5 e^2 N_f v_f^2 /\Delta$ (outside the range of the
 figure) for 
$\omega = 0.1 \Delta$.
The enhancement of the dissipative part at low frequencies is 
a consequence of the coupling of quasiparticles and 
collective order parameter modes, and cannot be obtained from
non-selfconsistent calculations.
Fig.\ref{condx}b shows that 
the real part of the conductivity becomes negative in the 
region of  dipolar backflow on the $y$-axis.
This leads locally to a negative
 time averaged
 power absorption, $<\vec{j}\cdot\vec{E}>_t \; <0$, and a
 corresponding  gain in energy,
which  is compensated 
by the strongly enhanced dissipation of energy 
in the center of the vortex.
The dissipative part of the local conductivity
at  a distance $R$ from the vortex
center exhibits pronounced maxima whose frequencies increase with 
increasing $R$ and are given by  2$\times$ the energy at the 
maxima in the
local density of states shown in Fig.2a. Hence, these features
in the absorption spectrum must be identified as
impurity assisted transitions between corresponding bound
states at negative and positive energies. Impurities are
required for  breaking  angular momentum conservation in these
transitions.

The applied electric field $\delta\vec{E}(t)$
induces in the vortex core 
an internal  field $-\vec{\nabla}\delta\varphi(t)$,
which is of the same order as the applied field.
Fig.\ref{fpatt1} shows the total electric field, 
$\delta\vec{E}_{tot}(t)=\delta\vec{E}^\omega(t)
-\vec{\nabla}\delta\varphi(t)$, 
in the vortex core. 
The induced field is at low frequencies  of  dipolar form,
and  oscillates out of phase (phase shift $\pi/2$) with
the applied field. This dipolar field originates from  small
charge fluctuation in the vortex core.
At higher frequencies the dipolar
field oscillates with a phase shift of $\approx \pi$, and 
screens part of  the applied field.

\begin{figure}[t]
%%%%%%%%%%%%%%%%%%%   F I G U R E   %%%%%%%%%%%%%%%%%%%%
\begin{minipage}[t]{6.0cm}
\centerline{
\epsfxsize3.8cm
\epsffile{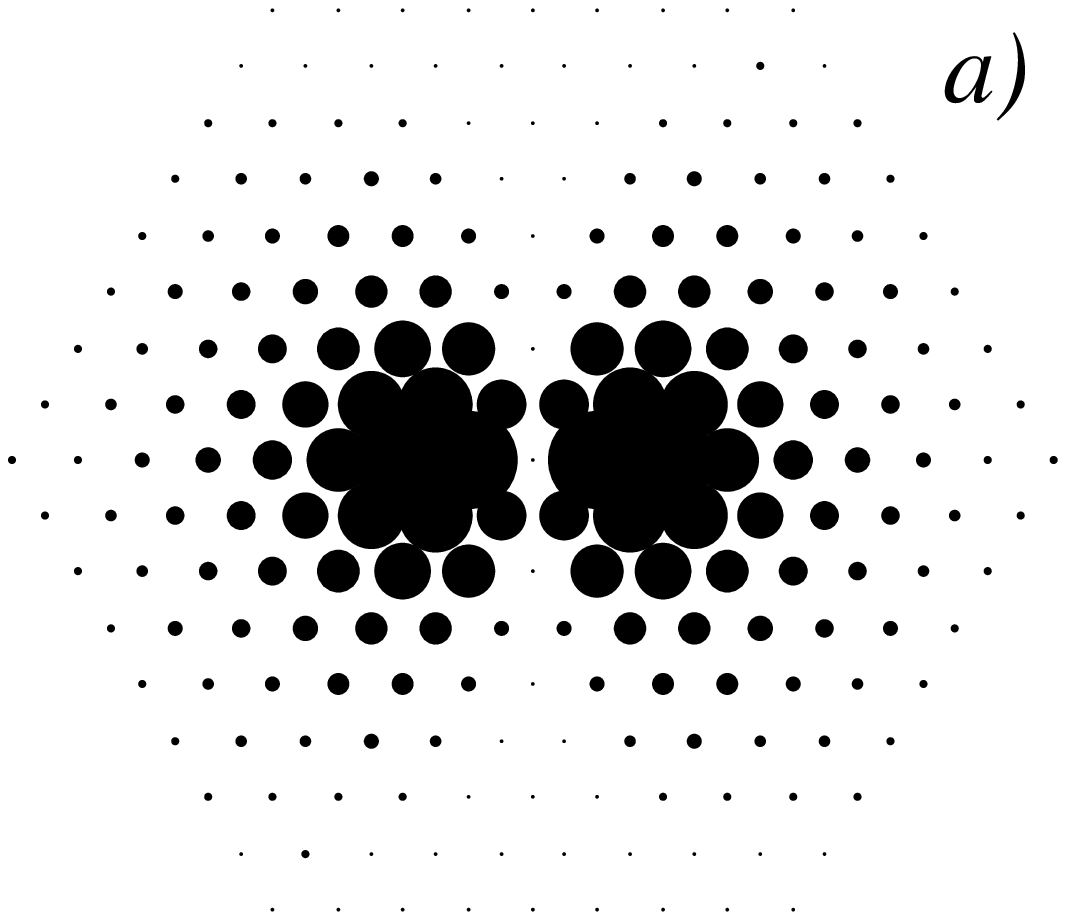}
}
\end{minipage}
\begin{minipage}[t]{6.0cm}
\centerline{
\epsfxsize3.8cm
\epsffile{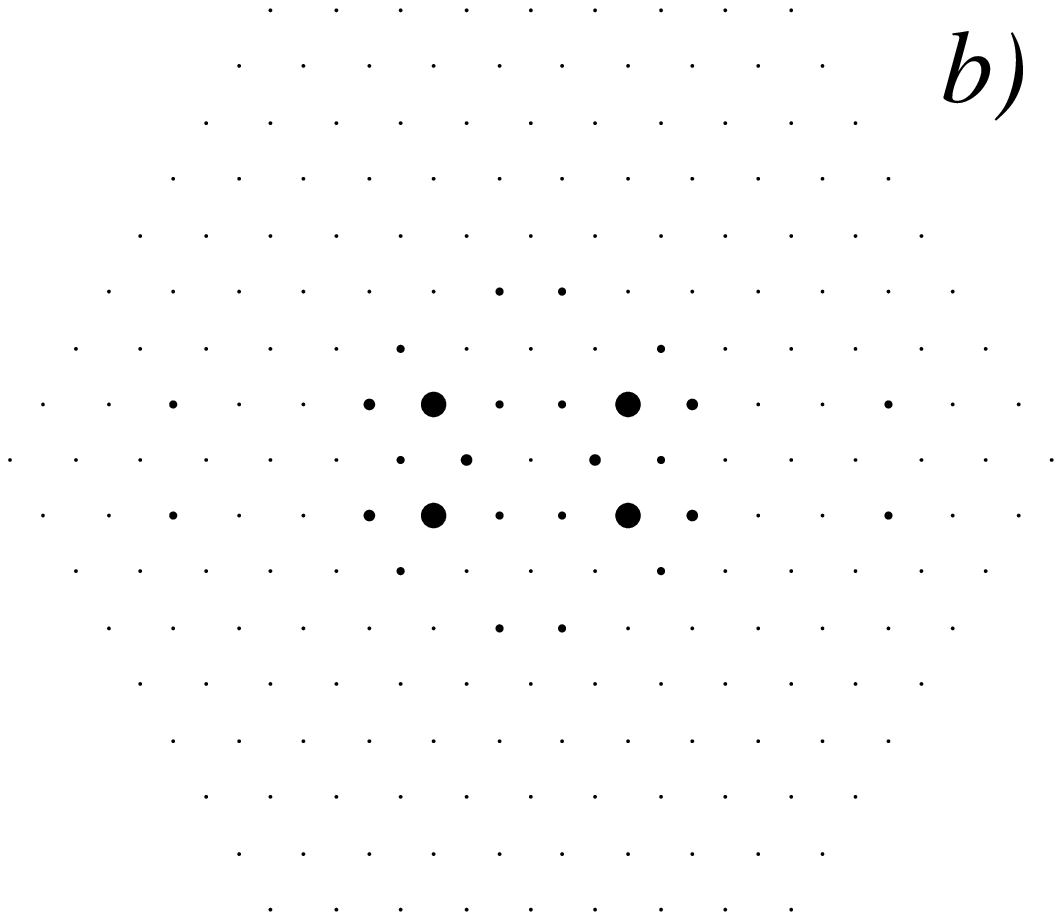}
}
\end{minipage}
\caption[]{ \label{diverg} 
Degree of violation of charge conservation by 
the dissipative  current flow
for a non-selfconsistent calculation
(Fig.\ref{diverg}a), and a   
a self-consistent calculation (Fig.\ref{diverg}b). 
The largest deviation in the non-selfconsistent calculation
amounts  to 
$\delta\dot{\rho}+\vec{\nabla} \delta\vec{j} = 
2.5 e^2 N_f v_f \delta E^\omega$.
The data are obtained for a superconductor with $\ell = 10
\xi_0 $ at $\omega = 0.4 \Delta (T)$ and  $T=0.3T_c$.}
%%%%%%%%%%%%%%%%%%%%%%%%%%%%%%%%%%%%%%%%%%%%%%%%%%%%%%%%
\end{figure}

We finally discuss the role of self-consistency in our calculation.
Our results were  obtained  by iterating the
self-consistency equations until the  relative error stabilized
below $\leq 10^{-10}$. Fig.\ref{diverg} compares the degree of
violation of
charge conservation  in a non-selfconsistent calculation
(no iteration) with the self-consistent result. 
We measure the degree of violation  at
position $\vec{R}$ by 
$D(\vec{R})=\max_{\, t}[ \delta\dot{\rho}(\vec{R},t) + 
\vec{\nabla}\cdot \delta \vec{j}(\vec{R},t)]$.
Charge conservation is obviously 
fulfilled if $D(\vec{R})=0$.
The degree of violation at a point $\vec{R}$ 
is indicated in Fig.\ref{diverg}a,b
by the size of the filled
circles around the grid points. The non-selfconsistent
calculation (Fig.\ref{diverg}a) results in a 
 $D(\vec{R})$, 
 which is much larger than the time derivative of the
  correct charge density,
   $\delta\dot{\rho}(\vec{R},t)$, obtained from a self-consistent
   calculation with 
   $\delta \varphi (\vec{R},t)=0$ instead of charge neutrality condition.
In Fig.\ref{diverg}b we show the violation of charge
conservation for
our  self-consistent calculation.
The small remaining  $D(\vec{R})$
is here a consequence of   the finite grid
size used in  our calculations. 

\section{Discussion}
This paper presented  self-consistent calculations
of the electromagnetic response of a pancake vortex in  a
superconductor with  finite but long mean free path. 
This complements previous calculations 
 for perfectly clean systems,
 which were done self-consistently in the limit 
$\omega\rightarrow 0$
(\cite{Hsu93}), and  at finite frequencies
without a self-consistent determination of
the order parameter 
(\cite{Janko92}, \cite{Zhu93}). 

The
frequency range of interest in our calculations is of the order of 
the gap frequency, $\hbar\omega=\Delta$.
We have shown that at low frequencies 
($\hbar\omega<0.5\Delta$) the electromagnetic 
dissipation is strongly 
enhanced in the vortex cores above its normal state value,  
 and that this 
effect is a consequence  of the coupled dynamics of
low-energy quasiparticles excitations bound to the vortex core
 and collective order parameter modes.
The induced current density has  at  low frequencies 
a dipole-like behaviour, which results from  
an oscillating   motion   of the vortex perpendicular to
direction of the driving {\it a.c.} field. 
 The response of the vortex in the 
intermediate frequency range,
$.5\Delta\la\hbar\omega \la 2\Delta$,
is dominated by   bound states in the vortex core. We find  
peaks in the local dissipation  at twice the bound state energies.
At higher frequencies, $\hbar\omega > 2\Delta$, the 
conductivity approaches that of a  very clean homogeneous
superconductor, which  is  in good
approximation given by    the non-dissipative response of
an ideal conductor.

%
% ---- End of Document ----
%
\end{document}